\documentclass[aps,twocolumn,float,superscriptaddress,prd]{revtex4-1}
\usepackage{graphicx, epsfig, bm, amsmath,amssymb}
\usepackage{color}
\usepackage{hyperref}
\usepackage{wasysym}
\usepackage{array}

\definecolor{darkgreen}{cmyk}{0.85,0.2,1.00,0.35} 
\definecolor{purple}{cmyk}{0.5,1.0,0,0} 
\definecolor{darkblue}{cmyk}{1.0,1.0,0,0}

\newcommand{\curv}{{\cal R}}
\newcommand{\R}{\mathcal{R}}

\newcommand{\ep}{\epsilon_H}

\newcommand{\cs}{c_s}
\newcommand{\esq}{\left(}  
\newcommand{\dir}{\right)} 

\hypersetup{
   colorlinks = true,
   allcolors = darkblue}

\begin{document}

\title{Polarization Predictions for Inflationary CMB Power Spectrum Features}

\author{  Vin\'icius Miranda }
\affiliation{Department of Astronomy \& Astrophysics, Kavli Institute for Cosmological Physics,  Enrico Fermi Institute, University of Chicago, Chicago IL 60637}

\author{Wayne Hu}
\affiliation{Department of Astronomy \& Astrophysics, Kavli Institute for Cosmological Physics,  Enrico Fermi Institute, University of Chicago, Chicago IL 60637}

\author{ Cora Dvorkin}
\altaffiliation{Hubble Fellow}
\affiliation{Institute for Theory and Computation, Harvard-Smithsonian Center for Astrophysics, 60 Garden St.; Cambridge, MA 02138}

\begin{abstract}
We conduct a model-independent analysis of temporal features during inflation in
the  large-scale CMB temperature power spectrum allowing for the possibility of non-negligible
tensor contributions.   Of  20 principal components of the inflationary history,
 the  suppression of power  at low  multipoles beginning
with a glitch at  multipoles  $\ell \sim 20-40$ 
 implies deviations in 2-3 of them with 2-3$\sigma$ deviations in each, with larger values reflecting
cases where tensors are allowed.
If tensors are absent, the corresponding $E$-mode polarization features follow
a similar pattern but are predicted to be up to 
twice as large.   They offer the opportunity to soon double the significance of inflationary 
features or eliminate them as an explanation of temperature features.
The tensor degeneracy with features in the temperature power spectrum is broken not only by $B$ but also  by $E$-polarization.   A precision measurement
of $E$-mode polarization at multipoles from $\ell\sim 20-60$ can potentially provide
an independent constraint on tensors that is less subject to dust foreground uncertainties.
\end{abstract}

\maketitle

\section{Introduction}

Given the soon to be released measurements of the large-angle polarization power spectrum, 
it is timely to reassess the status of anomalies in the temperature power spectrum and
determine how their inflationary origin might be confirmed or refuted with polarization
measurements.
Ever since the first release of WMAP data \cite{Hinshaw:2003ex}, the large-angle
temperature power spectrum has shown several anomalous features when compared
with the simplest power law or scale-free inflationary $\Lambda$CDM model.   In particular,
there is a glitch in the power spectrum at multipoles $\ell \sim 20-40$ \cite{Peiris:2003ff}
and a deficit
of large-angle correlations \cite{Spergel:2003cb,Copi:2006tu}.
%
%

The significance and interpretation of these features change with temperature measurements
by the Planck satellite \cite{Ade:2013nlj} and the 150 GHz 
measurement of degree scale $B$-mode 
polarization by the BICEP2 experiment  \cite{Ade:2014xna}.  Relative to new Planck data
at higher multipole moments,  the significance of the 
power deficit at low multipoles in the $\Lambda$CDM model increases.    Although the BICEP2
measurement is expected to be at least partially contaminated by galactic dust based on subsequent
Planck measurements at dust dominated frequencies \cite{Adam:2014bub}, any contribution from inflationary
gravitational waves near a tensor-to-scalar ratio of $r\sim 0.1$  \cite{Ade2013}
exacerbates the power spectrum deficit and increases its significance.

These temperature power spectrum features could indicate features in the initial spectrum
of curvature fluctuations.   Indeed there is an extensive literature on converting
these measurements into model-independent constraints on this spectrum
(e.g.~\cite{Hannestad:2000pm,Hu:2003vp,Tegmark:2002cy,Hannestad:2003zs,Bridle:2003sa,Mukherjee:2003ag,Leach:2005av,PhysRevD.81.021302,Hlozek:2011pc,Gauthier:2012aq,Vazquez:2012ux,Hunt:2013bha,Aslanyan:2014mqa,Hazra:2014jwa}).
Features
could even have an  origin in inflation if its
near time-translation invariance is broken at least transiently when these scales
left the horizon during inflation (e.g.~\cite{Peiris:2003ff,Contaldi:2003zv,Martin:2003sg,Freivogel:2005vv,Covi:2006ci,Joy:2007na,Hazra:2010ve,Achucarro:2013cva,Contaldi:2014zua,Miranda:2014wga,Abazajian:2014tqa,Hazra:2014jka,Bousso:2014jca}). 
Specific models that fit features also make predictions for $E$ mode polarization by which
they can be verified \cite{Mortonson2009,Miranda:2014wga,Bousso:2014jca}.   Here we seek to generalize these results and polarization predictions for any
single-field inflation model that fits the temperature features.   

Model independent constraints on curvature power spectrum features cannot be
directly applied to inflationary features.    In particular, not all possible curvature
power spectra are allowed in single field inflation.    This restriction is especially important
when considering sharp features in the temperature power spectrum.   
A temporal feature that is localized to less than an efold during  inflation does not
produce a feature in the curvature spectrum localized to a comparable range in
wavenumber as implied by the slow-roll approximation.   Instead the features
oscillate or ring across an extended range in wavenumbers.

For models with inflationary features, the  slow-roll approximation must
be replaced by either an exact computation or the  generalized slow-roll (GSR) approximation \cite{Stewart2002,Choe2004,Dvorkin2010}.    The latter has the advantage that
the curvature power spectrum depends on integrals which are linear in a single
source function.    Therefore, it is well suited for model-independent studies of power spectrum reconstruction \cite{DvoHu10a,Dvorkin:2011ui}.
In canonical single-field inflation, this source function is related to the shape of the inflationary potential in the same way the tilt is in the slow-roll approximation. 
In this paper, we adapt this reconstruction technique for the study of large-angle
power spectrum anomalies in the presence of potentially non-negligible tensor
contributions from inflation.

 We  start by reviewing and adapting the GSR formalism for inflationary source reconstruction with
 large-angle temperature data
 in \S\ref{sec:recon}. In \S\ref{sec:results}, we present results for the implied
 inflationary features and how their existence may be verified or falsified by $E$-mode
 polarization data.    We discuss these results in \S\ref{sec:discussion}.

\vfill
\section{Inflationary Reconstruction}
 \label{sec:recon}
 
 In this section we describe the GSR parameterization of inflationary sources to the curvature
spectrum.   We adapt the methodology of Refs.~\cite{DvoHu10a,Dvorkin:2011ui} for a focused study of large-angle
anomalies including the possibility of non-negligible tensor contributions.   
In \S \ref{sec:GSR} we review the GSR formalism itself and its application to the curvature
and tensor power spectra.   We describe the sampling parameters for the
curvature source used in the likelihood analysis as well as representations in terms
of principal components in \S \ref{sec:source}.

\subsection{Generalized Slow Roll}
\label{sec:GSR}

The GSR approach provides a model-independent description of inflationary
power spectra that allows for transient violations of the ordinary slow-roll approximation in
single field inflation
as might occur from features in the inflaton potential or sound speed evolution.
In this approximation, 
features in the curvature power spectrum arise from changes in the Hubble rate $H$
and sound speed $c_s$ through a single function of time or scale
\begin{equation}
G (\ln s)= - 2 \ln f    + {2 \over 3} (\ln f )'   ,
\end{equation}
where \cite{Stewart:2001cd,Hu:2011vr}
\begin{align}\label{eqn:fdef}
	f^2 & = 8 \pi^2 \frac{\ep \cs}{H^2} \esq \frac{a H s}{\cs} \dir^2.
\end{align}
Here,
\begin{equation} \label{eqn:acceleration}
\epsilon_H=-\frac{d\ln H}{dN}
\end{equation}
where $N$ is the number of efolds and $N=0$ at the end of inflation,
$c_s$ denotes the sound speed of field fluctuations, and $' = d/d\ln s$. The sound horizon is given by
\begin{align}\label{eqn:def_sound_horizon}
	s(N) = \int_{N}^{0} d \tilde{N} {c_s \over aH}.
\end{align}
Specifically the dimensionless curvature power spectrum as derived from Green function 
techniques is
  \cite{Choe:2004zg,Dvorkin:2009ne}
\begin{align} \label{eqn:GSRpower}
\ln \Delta_\curv^{2}(k) &\approx  G(\ln s_{*}) + \int_{s_{*}}^\infty {d s\over s} W(ks) G'(\ln s)\\
&\quad + \ln \left[ 1+ I_1^2(k) \right], \nonumber
\end{align}
where $s_*$ is an arbitrary epoch during inflation such that all relevant $k$-modes
are well outside the sound horizon, $k s_* \ll 1$.
Changes in the source function $G'$ are transferred to the power spectrum according
to the window function
\begin{align}
	\label{eqn:powerwindow}
W(x) &= {3 \sin(2 x) \over 2 x^3} - {3 \cos (2 x) \over x^2} - {3 \sin(2 x)\over 2 x},
\end{align}
at leading order in the deviations of the inflaton mode function from its de Sitter form and to 
\begin{align}
X(x) & = {3 \over x^3} (\sin x - x \cos x)^2 ,
\end{align}
through
\begin{eqnarray}
I_1(k) &=& { 1\over \sqrt{2} } \int_0^\infty {d s \over s} G'(\ln s) X(ks),  
\end{eqnarray}
for the dominant second order contribution.   This form for the power spectrum remains a good approximation
if the second order term \cite{Dvorkin:2011ui}
\begin{equation}
I_1 < \frac{1}{\sqrt{2}} ,
\label{eqn:I1criterion}
\end{equation}
and hence allows for up to order unity features in the curvature power spectrum.

Given the dependence of this power spectrum on a single source function, we seek to
constrain or reconstruct $G'(\ln s)$ directly from the data.
Note that for an inflaton with a canonical $c_s=1$ kinetic term \cite{Dvorkin:2009ne}
\begin{equation}
G' \approx 3 \left( \frac{V_{,\phi}}{V} \right)^2 - 2 \frac{V_{,\phi\phi}}{V},
\label{eqn:potential}
\end{equation}
so that a reconstruction of $G'$ can be thought of as a measurement of the shape
of the potential.   More generally in $P(X,\phi)$ inflation \cite{Hu2011}, or equivalently the effective field theory of inflation for the $\pi$ mode of the broken time translation invariance \cite{Cheung:2007st}, 
$G'$ is the quantity that determines the tilt in the ordinary slow-roll approximation.

In principle, the gravitational wave power spectrum in each polarization state $\Delta^2_{+,\times}$ follows the same prescription with
the replacement \cite{Gong:2004kd,Hu:2014hoa}
\begin{equation}
f^2 \rightarrow f_h^2 =  \frac{ 2\pi^2}{H^2} (a H\eta)^2,
\end{equation}
where $\eta$ is the conformal time to the end of inflation.   However, since this source
only depends on $H$ and integrals of $H$, transient changes in $\epsilon_H$ and $c_s$
have very small impact on tensors.   The spectrum therefore remains a power law to good approximation, and we parameterize it as usual by a tensor-to-scalar ratio $r$ and tilt $n_t$,
\begin{equation}
\Delta^2_{+,\times}(k) = \frac{r}{4} \Delta^2_\curv(k_0) \left( \frac {k}{k_0} \right)^{n_t},
\end{equation}
where $k_0$ is some fiducial normalization scale.   If $k_0$ is chosen to be far from any features
in the curvature power spectrum then the usual consistency relation $n_t= -r/8c_s$ applies.

\subsection{Parameterized Source}
\label{sec:source}

The ordinary slow-roll approximation corresponds to a parameterization of the
curvature source function by  a constant $G'(\ln s) = 1-n_s$, and results in 
a power-law curvature power spectrum.   We therefore  look for
parameterized deviations from this constant behavior.
In general, given some set of
basis functions $B_i(\ln s)$ we can describe the source function with a set
of coefficients $p_i$ as
\begin{align}
       \delta G'(\ln s) &\equiv G'(\ln s) - (1-n_s) \nonumber\\
	&=  \sum_i p_i B_i(\ln s).
	\label{eqn:Gpbasis}
\end{align}
The advantage of the GSR form in Eq.~(\ref{eqn:GSRpower}) is that the integrals
are linear in $G'$ and hence the impact of the individual components can 
be precomputed separately 
\begin{align} \label{eqn:gsr_window_integrals}
W_i(k) &= \int_{s_*}^{\infty} {d s\over s} W(k s) B_i(\ln s) , \nonumber\\
X_i(k) &= \int_{0}^{\infty}  {d s\over s} X(k s) B_i(\ln s),
\end{align}
so that the power spectrum becomes a sum over the basis
\begin{align} \label{eqn:ps2v2_basis}
	\ln \Delta_{\mathcal{R}}^2(k) =&  \ln A_s\left(\frac{k}{k_0}\right)^{n_s-1} 	+ \sum_i p_i \big[ W_i(k) - W_i(k_0) \big] \nonumber\\
	&+ \ln \left[ \frac{1+ I_1^2(k)}{1+ I_1^2(k_0)} \right],
\end{align}
where
\begin{align}
I_1(k) =  \frac{ \pi}{2\sqrt{2}}(1-n_s) +\frac{1}{\sqrt{2}} \sum_i p_i X_i(k).
\end{align}
Note that we have absorbed the normalization constant $G(\ln s_*)$ into the
amplitude of the power spectrum at the scale $k_0$
\begin{align}
A_s =  \Delta_{\R}^2(k_0).
\end{align}

In Ref.~\cite{DvoHu10a,Dvorkin:2011ui}, the basis functions $B_i$ were chosen to be
the principal components (PCs) of the Fisher matrix for the full WMAP range of scales. 
Since the Fisher matrix is constructed from the expected errors of a given experiment,
this technique is blind to the presence of anomalies in the actual data.    The drawback
for studying known large-angle anomalies is that the basis does not efficiently encode them.
Here we take an alternate approach that is better suited to making polarization predictions
for such anomalies rather than searching for them.

These anomalies appear on scales larger than the acoustic scale at recombination but smaller
than the current horizon scale, and so we choose to restrict our parameterization  to
\begin{equation}
200 < \frac{s}{\rm Mpc}< 20000.
\label{eqn:range}
\end{equation}
Next we follow Ref.~\cite{Dvorkin:2011ui} in defining a band limit for the frequency of
deviations by sampling $\delta G'(\ln s_j)$ at a rate of 10 per decade or about 4 per efold
of inflation.  This rate was determined to be sufficient to capture large-scale features
in the power spectrum.  

The parameterized $\delta G'$ function is then the natural spline of these sampling points
$p_i = \delta G'(\ln s_i)$.
In the $B_i$ language of Eq.~(\ref{eqn:Gpbasis}), its basis is constructed 
by splining the set of sampling points 
\begin{equation}
B_i(\ln s_j) = \begin{cases}
1 & i=j \\
0 & i\ne j \\
\end{cases},
\end{equation}
with $\ln s_i$ values in the range specified by Eq.~(\ref{eqn:range}) and a sampling grid in $\ln s_j$ that
extends sufficiently further that the basis functions have negligible support thereafter.
We choose the arbitrary $\ln s_*$ epoch to be the large-scale endpoint of the sampling grid and order the points so that $s_1=200$ Mpc is the smallest scale.

In practice, we then precompute $W_i(k)$ and $X_i(k)$ on a fine grid in $k$-space
and use a modified version of CAMB to evaluate CMB observables.  
The curvature power spectrum is then defined by 22 parameters
$\{ \ln A_s, n_s, p_1, \ldots p_{20} \}$.    We set the normalization point
$k_0=0.08$ \,Mpc$^{-1}$ to be  the pivot point for the Planck dataset \cite{Miranda:2013wxa}, which has the benefit that it 
is in the featureless or slow-roll regime by assumption.  
To these we add the cosmological parameters of the 
flat $\Lambda$CDM model: the cold dark matter density
$\Omega_c h^2$, the baryon density $\Omega_b h^2$, the effective angular size of the CMB 
sound horizon $\theta_{\rm MC}$, and the Thomson optical depth to recombination $\tau$.  
We call this model  $G\Lambda$CDM, defined by 26 parameters, whereas the $\Lambda$CDM model sets $p_i=0$ and has only 6 free parameters.

For the tensor power spectrum we consider
cases where $r=0$ or constrained by the temperature and/or polarization data.   Since there is little current information on the
slope of the tensor spectrum, we set $n_t= -r/8$ so as to satisfy the inflationary
consistency relation for $c_s=1$.     We call the model that allows for
non-negligible tensors $rG\Lambda$CDM.

Constraints on these parameters from the
datasets are obtained using the Markov Chain Monte Carlo (MCMC) technique implemented with the CosmoMC
code \cite{Lewis:2002ah}.   The cosmological parameters are all given non-informative priors except for a global constraint
on $I_1$ set by Eq.~(\ref{eqn:I1criterion})
beyond which the GSR approach breaks down.   We also include the standard Planck
foreground parameters in all analyses.

Since our choice of  parameters oversamples $\delta G'$ relative to what the
data can constrain, individual measurements of $p_i$ are noisy with any true signal
buried in the small covariance between parameters.    For visualization purposes, we therefore also construct the principal 
components derived from an eigenvalue decomposition of the MCMC covariance matrix estimate
\begin{align}
C_{ij} &= \langle p_i p_j \rangle - \langle p_i \rangle \langle p_j \rangle \nonumber\\
 &= \sum_a S_{ia} \sigma_a^2 S_{ja},
\end{align}
where $S_{ia}$ is an orthonormal matrix of  eigenvectors.   Specifically, we define 
the PC parameters 
\begin{align}
m_a = \sum_i S_{ia} p_i,
\end{align}
such that their covariance matrix satisfies
\begin{align}
\langle m_a m_b \rangle - \langle m_a \rangle \langle m_b \rangle = \delta_{a b} \sigma_a^2.
\end{align}
We then postprocess the MCMC chains to obtain the posterior probability distributions in
these derived parameters.
Given a rank ordering of the PC modes from smallest to largest variance, we can
also construct a PC filtered reconstruction of $\delta G'$ as \cite{Dvorkin:2011ui}
\begin{equation}
\delta G'_{b\rm PC}(\ln s_i) = \sum_{a=1}^{b} m_a S_{ia},
\label{eqn:PCfilter}
\end{equation}
where $b$ is chosen to reflect the well-measured eigenmodes.

The differences between this construction and that of Ref.~\cite{Dvorkin:2011ui} are that
the PCs are defined by the covariance matrix inferred from the data itself and change for
different data combinations, their normalization is
set by the discrete rather than continuous orthonormality condition, and that their
range is restricted by Eq.~(\ref{eqn:range}) to be in the region of known anomalies.
Finally, we allow for the possibility of non-negligible tensor contributions to the observed
spectra through the tensor-to-scalar ratio $r$.

\begin{table}[t] \centering
\def\arraystretch{1.40}
\begin{tabular}{| c |  c | c | }
\hline  
& Model & Dataset \\ \hline
$G$-$T$ & $G\Lambda$CDM & $T$ \\ 
$rG$-$T$& $rG\Lambda$CDM & $T$ \\
$rG$-$TB$ & $rG\Lambda$CDM & $T$+BICEP2\\
 \hline
\end{tabular}
\caption { \footnotesize Models and datasets.  The $G\Lambda$CDM model includes 20 parameters that sample  curvature source function deviations in addition to the 6 flat power-law $\Lambda$CDM parameters, whereas the  $rG\Lambda$CDM includes the 
tensor-to-scalar ratio $r$.   The $T$ dataset mainly reflects Planck temperature data, but also includes WMAP9 polarization, Union 2.1 supernovae distance,
baryon acoustic oscillation, and $H_0$ measurements to constrain cosmological parameters.   The BICEP2 data
set adds polarization constraints that limit $r$.
}  
 \label{table:data_sets} 
\end{table}

\begin{table*}[t] \centering
\def\arraystretch{1.40}
\begin{tabular}{| c || c | c |  c | }
\hline  
 & $G$-$T$ & $rG$-$T$ & $rG$-$TB$ \\ \hline
 $\Omega_b h^2$        &  $0.02218\pm 0.00024$ &    $0.02210\pm 0.00025$ & $0.022093\pm 0.00024$ \\
 $\Omega_c h^2$        & $0.1183\pm 0.0014$      & $0.1183 \pm 0.0014$ & $0.1182\pm 0.0014$\\
 $\theta_{\text{MC}}$   & $1.04150\pm 0.00055$  & $1.04142 \pm 0.00054$& $1.04148\pm 0.00054$  \\  
 $\tau$                         & $0.099\pm 0.016$         & $0.100\pm 0.017$          & $0.103\pm 0.017$ \\
 $\ln (10^{10} A_s)$     & $3.086\pm 0.033$         & $3.089\pm 0.033$          & $3.094\pm 0.034$ \\
  $n_s$                        & $0.9612\pm 0.0060$      & $0.9638\pm 0.0063$      & $0.9626\pm 0.0060$\\
  $r$                             & $0$                                & $0.30\pm 0.16$              & $0.229\pm 0.048$\\
  \hline
  $p_1$                        & $-0.05\pm 0.10$            & $-0.17\pm 0.13$             & $-0.11\pm 0.11$ \\
  $p_2$                        & $-0.13\pm 0.17$            & $-0.20\pm 0.17$             & $-0.16\pm 0.18$\\
  $p_3$                        & $0.07\pm 0.24$             & $-0.16\pm 0.27$             & $-0.03\pm 0.24$\\
  $p_4$                        & $-0.47\pm 0.35$            & $-0.57\pm 0.35$             & $-0.60\pm 0.34$\\
  $p_5$                        & $0.51\pm 0.56$              & $0.37\pm 0.56$             & $0.41\pm 0.53$\\
  $p_6$                        & $-0.61 \pm 0.92$           & $-0.69\pm 0.94$             & $-0.84\pm 0.90$ \\
  $p_7$                        & $-0.8\pm 1.5$                & $-0.8\pm 1.5$                 & $-0.6\pm 1.4$ \\
  $p_8$                        & $-0.4\pm 2.5$                & $-0.5\pm 2.4$                 & $-0.8\pm 2.4$ \\
  $p_9$                        & $-0.2\pm 3.3$                & $0.1\pm 3.0$                  & $0.4\pm 3.0$ \\
  $p_{10}$                    & $2.2\pm 3.5$                 & $1.5\pm 3.1$                  & $1.3\pm 3.1$ \\
  $p_{11}$                    & $-0.4\pm 3.5$                & $-0.1\pm 3.3$                 & $0.1\pm 3.3$ \\
  $p_{12}$                    & $-1.3\pm 3.3$                & $-1.3\pm 3.3$                 & $-1.5\pm 3.2$ \\
  $p_{13}$                    & $-0.0\pm 3.2$                & $0.1\pm 3.3$                  & $0.2\pm 3.2$\\
  $p_{14}$                    & $1.2\pm 3.3$                 & $1.1\pm 3.5$                  & $1.0\pm 3.4$\\
  $p_{15}$                    & $-1.4\pm 3.6$                & $-1.2\pm 3.7$                 & $-1.1\pm 3.6$ \\
  $p_{16}$                    & $-0.1\pm 3.9$                & $-0.1\pm 3.9$                 & $-0.3\pm 3.9$ \\
  $p_{17}$                    & $0.3\pm 4.1$                 &  $0.2\pm 4.1$                 & $0.3\pm 4.1$ \\
  $p_{18}$                    & $-0.1\pm 3.9$                & $-0.1\pm 3.9$                 & $-0.2\pm 3.9$\\
  $p_{19}$                    & $0.04\pm 3.7$               & $0.0\pm 3.8$                  & $0.1 \pm 3.7$\\
  $p_{20}$                    & $0.1\pm 2.9$                 & $0.0 \pm 2.9$                 & $0.0\pm 2.9$ \\\hline
  
  $m_1$ & $-0.155\pm 0.080$ & $  -0.018\pm 0.095$ & $-0.188\pm 0.084$  \\
  $m_2$ & $-0.26\pm  0.12$  & $ -0.27\pm 0.16$ & $-0.41\pm 0.12$ \\
  $m_3$ & $-0.22\pm 0.18$ & $-0.65\pm 0.20$ & $ -0.40\pm 0.18$  \\\hline
\end{tabular}
\caption { \footnotesize Parameter constraints (68\% CL) for the various model-dataset combinations of 
Tab.~\ref{table:data_sets}.    $p_i$ represent the deviations in the curvature source function
from a scale-free power law and the derived parameters $m_a$ represent amplitudes of the principal components of their covariance matrix, which are not the same between combinations.
}
 \label{table:mc_parameters} 
\end{table*}

\section{Results}
\label{sec:results}

Here we present results for the curvature source function $G'$, which controls deviations from power-law initial conditions 
and their polarization predictions.  We begin in \S \ref{sec:curvature} with the case
where tensor contributions are assumed to be negligible.  In \S \ref{sec:tensors} we study
the impact of tensors, constrained either by the temperature data alone or by the
BICEP2 $B$-mode measurement,
 in changing the interpretation of temperature anomalies and their
polarization predictions.   These model and dataset choices are summarized in
Tab.~\ref{table:data_sets}.

\begin{figure}[tbh]  
\psfig{file=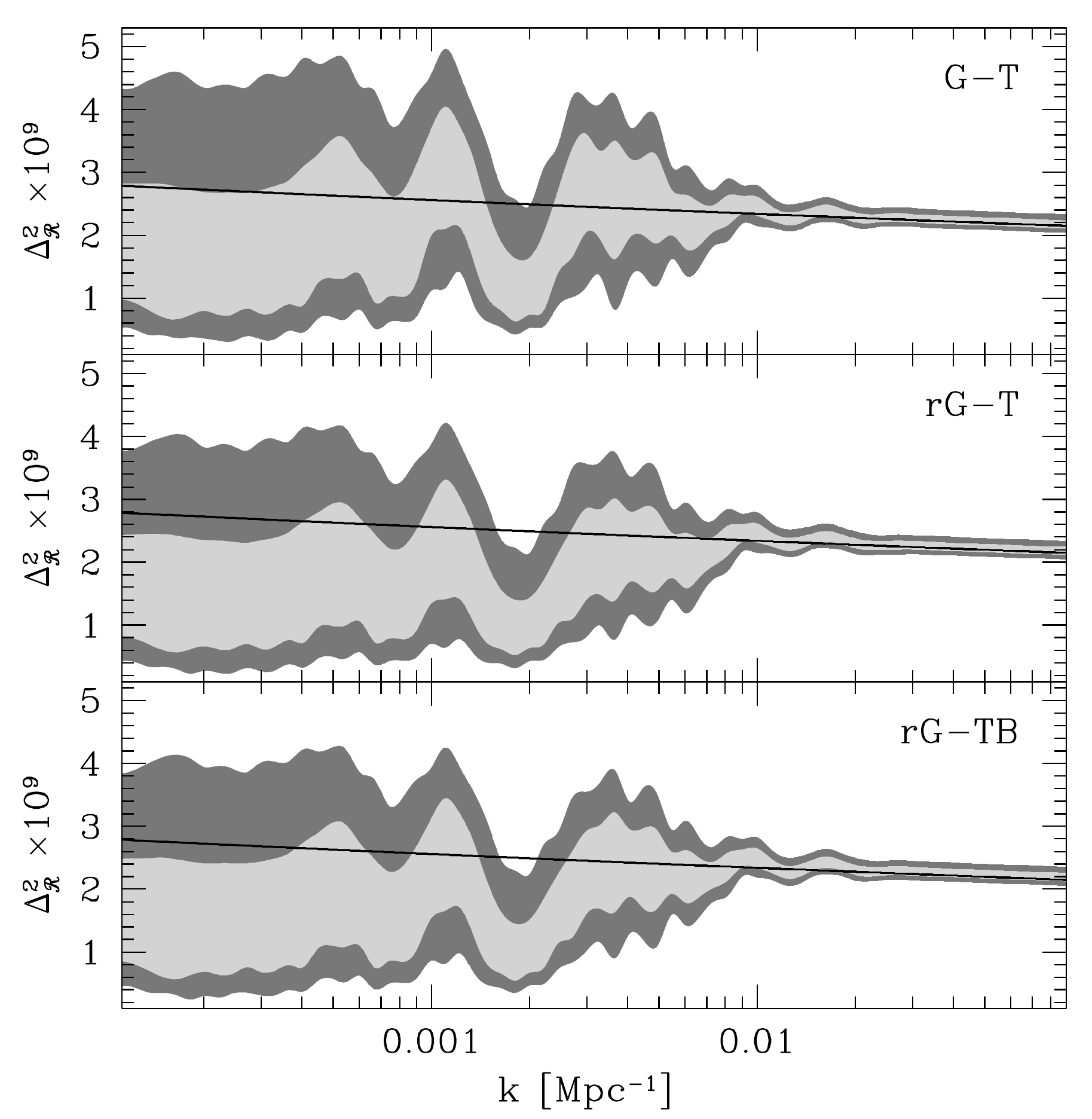, width=3.3in}
\caption{\footnotesize  Curvature power spectrum constraints derived
from those on the curvature source function in the various model-dataset combinations
(68\% and 95\% CL bands here and below).  With no tensors (top panel) the suppression of power at $k \lesssim 0.002$\,Mpc$^{-1}$ begins at a sharp glitch with slightly larger power on either side. Allowing tensors in the $T$ dataset (middle panel) absorbs the excess at high $k$, decreasing the significance of the glitch but increasing that of the power suppression.   Constraining the
maximum allowed tensors in the $TB$ dataset (bottom panel) interpolates between these cases.  Lines represent the fiducial $\Lambda$CDM model which we use in the following figures as a baseline for comparison.}.
\label{fig:curvature}	
\end{figure}

\begin{figure}[tbh]  
\psfig{file=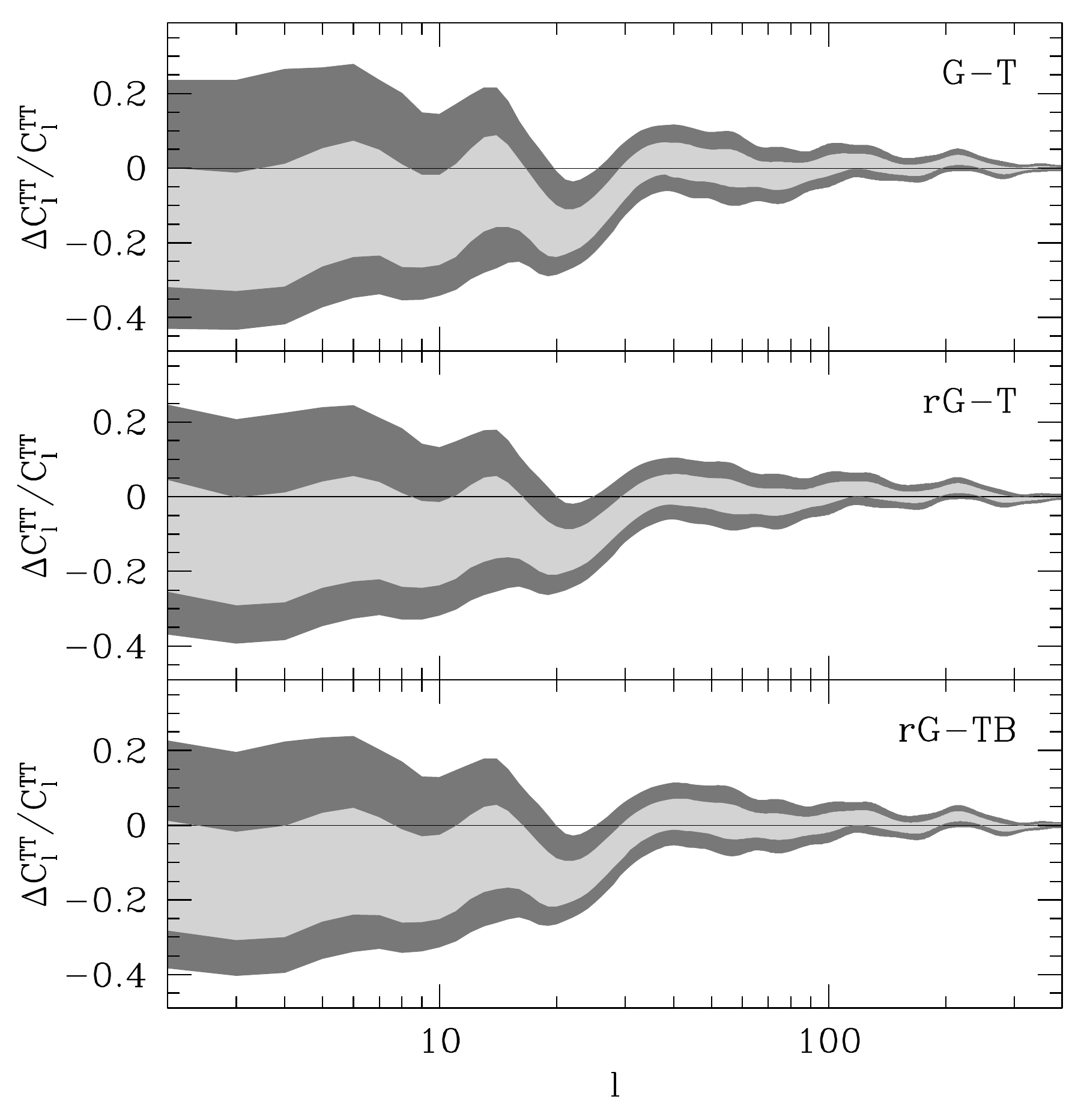, width=3.3in}
\caption{\footnotesize Temperature power spectrum constraints relative to the fiducial $\Lambda$CDM model of Fig.~\ref{fig:curvature}. In each model-data case, deviations in the curvature source $\delta G'$ can model the $\sim 15\%$ glitch feature at
$\ell \sim 20-40$ and the suppression of low multipole power but with different contributions from tensors that lead to different predictions for polarization and curvature sources.}
\label{fig:temp}	
\end{figure} 

\begin{figure}[tbh]  
\psfig{file=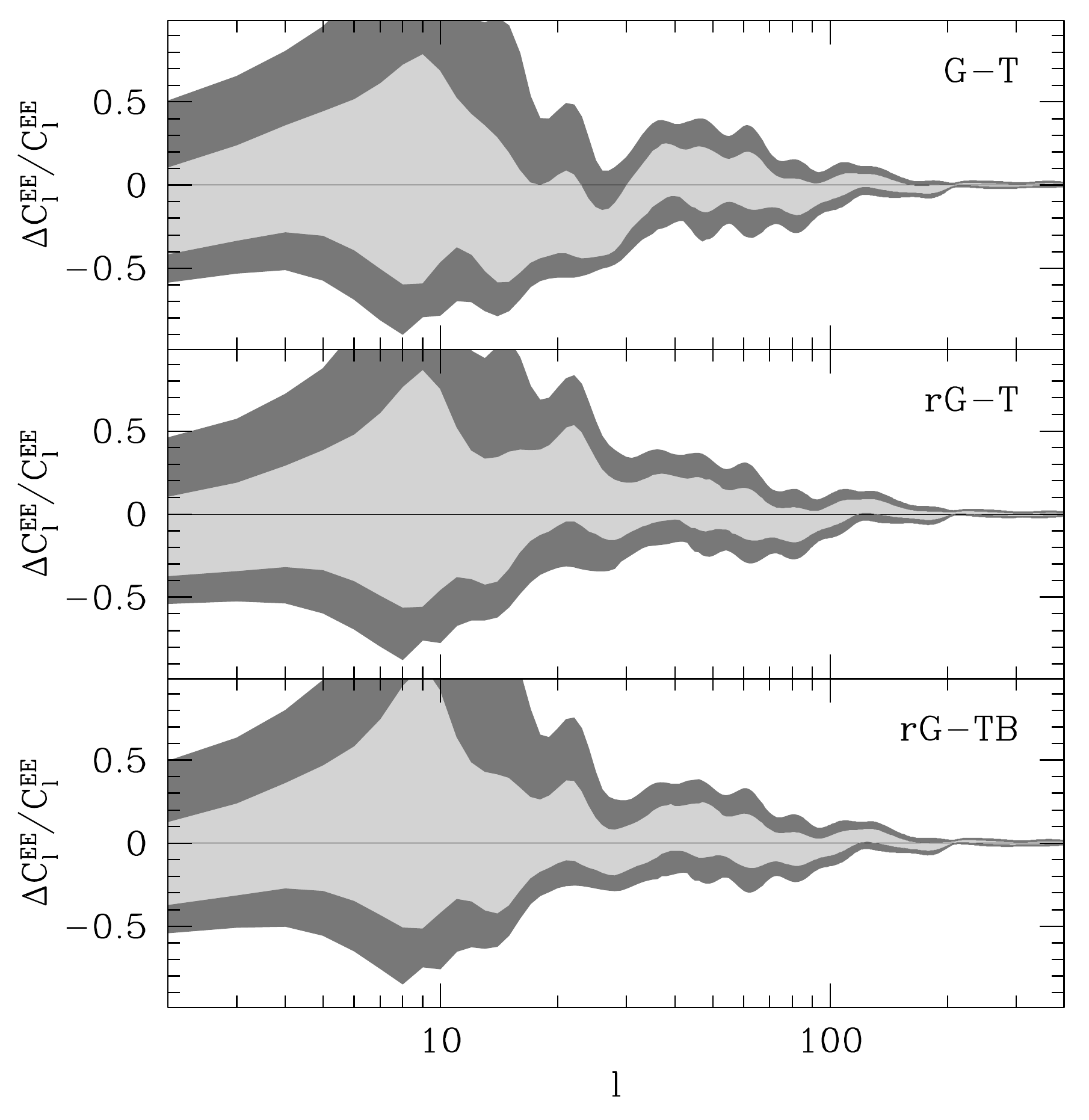, width=3.3in}
\caption{\footnotesize Polarization predictions for the various model-data combinations. With no tensors (top panel) predicted features are twice as large ($\sim 30\%$) as the corresponding temperature
ones, implying that comparably precise measurements should conclusively confirm or
falsify their origin as curvature source features.   With the tensors  allowed by the temperature based $T$ dataset (middle panel), the relatively larger tensor $E$ contributions fill in the
$\ell < 40$ scalar deficit with increments predicted for $r \gtrsim 0.2$.    Using the BICEP2 measurement to limit the $r$ bound, the possible increment and a measured decrement 
would provide independent constraints on $r$. 
Note that $\ell \lesssim 20$ predictions are subject to reionization model uncertainties and
employ WMAP9 polarization constraints that are subject to galactic foreground uncertainties. 
 }
\label{fig:pol}	
\end{figure}

\subsection{Curvature Only}
\label{sec:curvature} 

We begin with a baseline dataset whose inferences on the source function $G'$
 is mainly driven by the Planck temperature power spectrum \cite{Ade:2013zuv}.  To these we
 add the WMAP9 polarization \cite{Bennett:2012fp}, Union 2.1 supernovae distance
 \footnote{\href{http://www.supernova.lbl.gov/Union}{http://www.supernova.lbl.gov/Union}}, 
 baryon acoustic
 oscillation \cite{Anderson:2012sa, Padmanabhan:2012hf, Blake:2011en}, and SHOES $H_0$ 
 \cite{Riess:2011yx} datasets in order to constrain other cosmological
 parameters in the flat $\Lambda$CDM model.    We call this combination
 the ``$T$" dataset.  
 
 We first study this $T$ dataset under the assumption that tensors are 
 negligible ($r=0$) in the $G\Lambda$CDM context and call this the $G$-$T$ analysis.
 Tab.~\ref{table:mc_parameters} gives the constraints on parameters.    
 As expected, our oversampling of the $\delta G'$ function relative to what the data
 can constrain means that results on individual amplitudes $p_i=\delta G'(\ln s_i)$ 
 marginalized over the other parameters have very low signal-to-noise.  
 Nonetheless, combined with their covariances, they do favor a suppression of the curvature power
 spectrum at large scales.   To quantify these features we consider the power spectrum
 itself $\Delta_\curv^2(k)$ to be a derived parameter and show the 68\% and 95\%
 CL regions in Fig.~\ref{fig:curvature} (top panel).   
 We also show the best fit $\Lambda$CDM model \cite{Miranda:2013wxa}, with parameters  $\Omega_c h^2= 0.1200$, $\Omega_b h^2=0.02204$, $h=0.672$, $\tau=0.0895$, $A_s= 2.156 \times 10^{-9}$,  $n_s=0.961$, $r=0$ for reference.
   In the following we quote results
 relative to the predictions of this model.

 Note the coherent suppression of power relative to this fiducial model
 for $k \lesssim 0.002$ Mpc$^{-1}$ that begins with a fairly sharp, almost oscillatory, dip with a slight preference for larger power on either side.   
 This position 
 corresponds to the known temperature power spectrum anomaly at $\ell \sim 20-40$ as shown in Fig.~\ref{fig:temp} \cite{Peiris:2003ff}.    
 Here we similarly consider the theoretical temperature power spectrum 
 $C_\ell^{TT}$  as a derived parameter.
 In particular there is $\sim 15\%$ deficit of power at $\ell \lesssim 20$ and a slight excess of power at
 $\ell \sim 40$.

    In this region, the $TE$ cross correlation  is small
 and so the $EE$ power spectrum provides nearly independent information on this feature.
 In Fig.~\ref{fig:pol}, we show the theoretical $C_\ell^{EE}$ power spectrum as a 
 derived parameter.     Note first the much larger allowed range of fractional deviations.  
 Since the cosmic variance limit on measuring deviations in $TT$ and $EE$ are the same fractionally,
 this indicates the large discovery potential for precision $EE$ measurements with
 even $40\%$ measurements across the $\ell \sim 20-40$ band being of interest
 for verifying or falsifying the inflationary explanation of temperature features.  
Because of projection effects, namely the enhanced sharpness of the transfer of power 
to polarization \cite{Hu:1997hp},  the fractional suppression of polarization power is predicted to begin at a slightly higher
multipole and is allowed to reach lower values at the extrema at around $\ell\sim 26$.
 For $\ell \lesssim 20$ the predictions are subject to uncertainties in reionization 
 \cite{Mortonson2009} as
 well as possible impact of galactic foregrounds on the WMAP9 polarization used here
 as a constraint.   They are 
thus of less immediate relevance for inflationary features.

\begin{figure}[h]  
\psfig{file=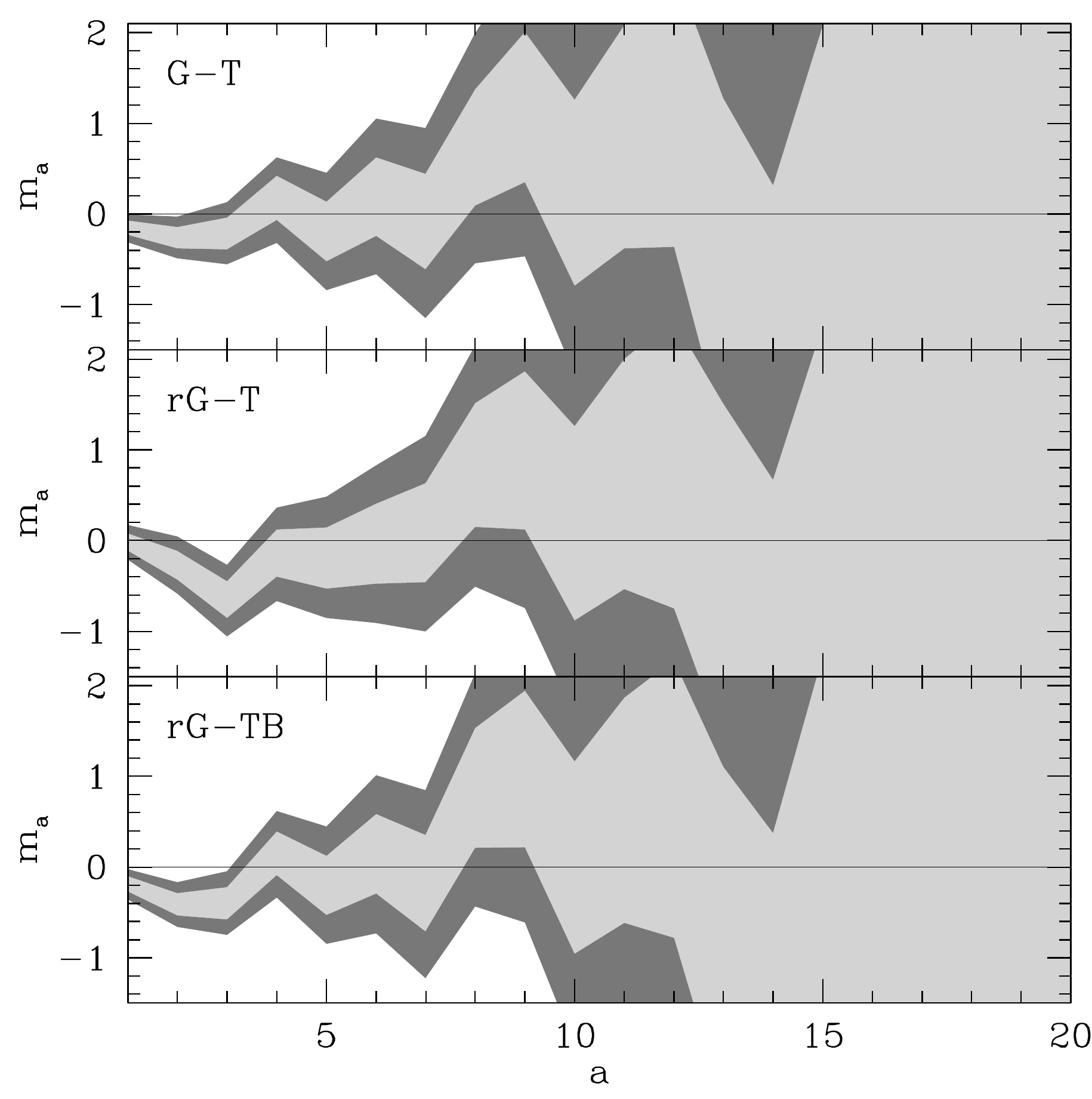, width=3.3in}
\caption{\footnotesize Principal component amplitude constraints for the curvature source function
$\delta G'$.   Deviations from a featureless $m_a=0$ spectrum at $>95\%$CL appear in the first
3 PCs but are absent in the higher ones.  Models with allowed tensor contributions show both
larger and more significant deviations.   The 3 PCs are constructed separately in each 
model-dataset combination and hence $m_a$ does not represent the same parameter
between panels. }
\label{fig:pc}	
\end{figure} 

\begin{figure}[h]  
\psfig{file=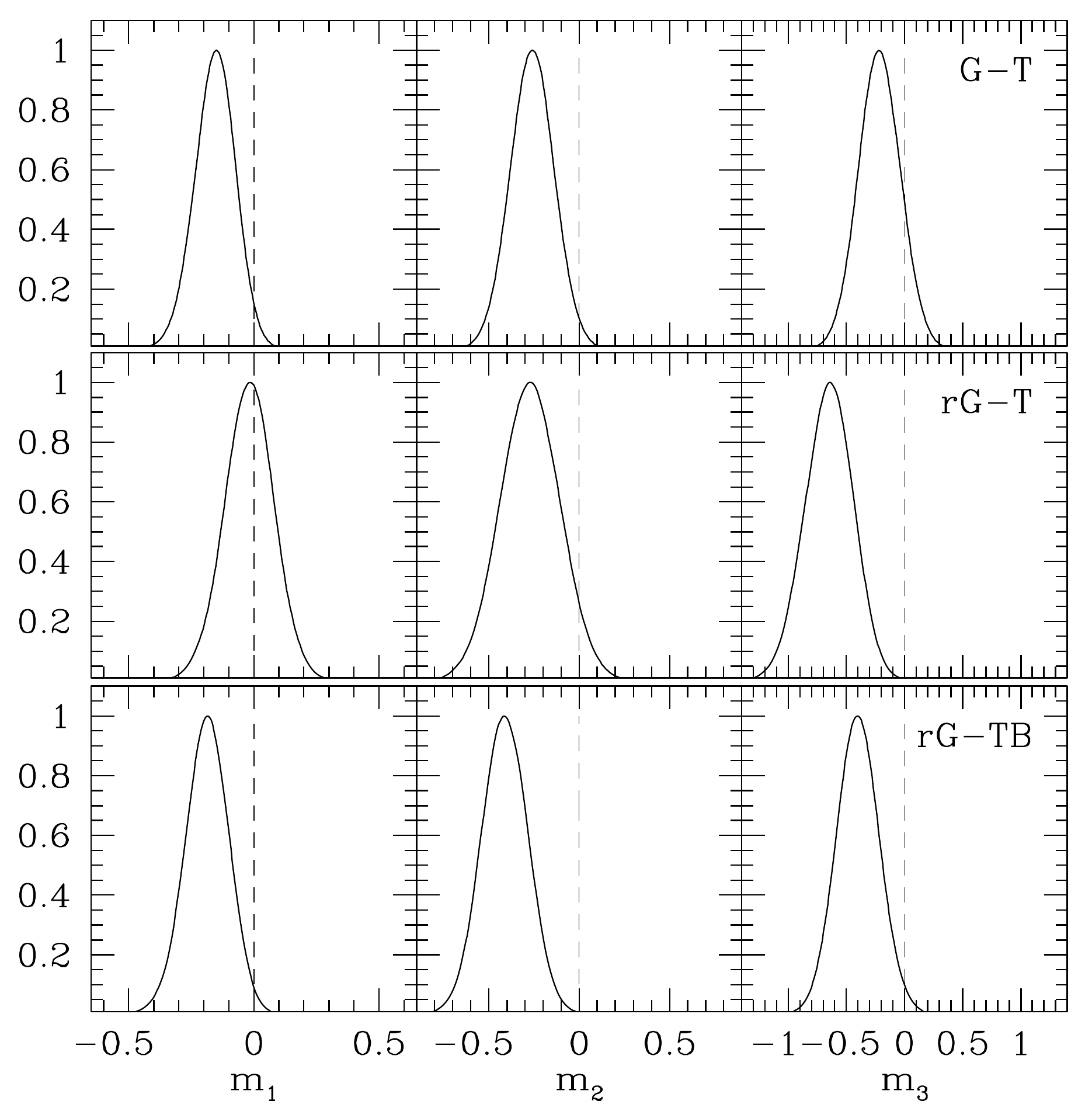, width=3.3in}
\caption{\footnotesize Posterior probability distributions of the first 3 PC parameters.  In each model-dataset
case, the featureless $m_a=0$ model lies in the tails for two or three components, 
with the more extreme deviations for those that allow tensor contributions.}
\label{fig:PCpost}	
\end{figure} 

\begin{figure}[h]  
\psfig{file=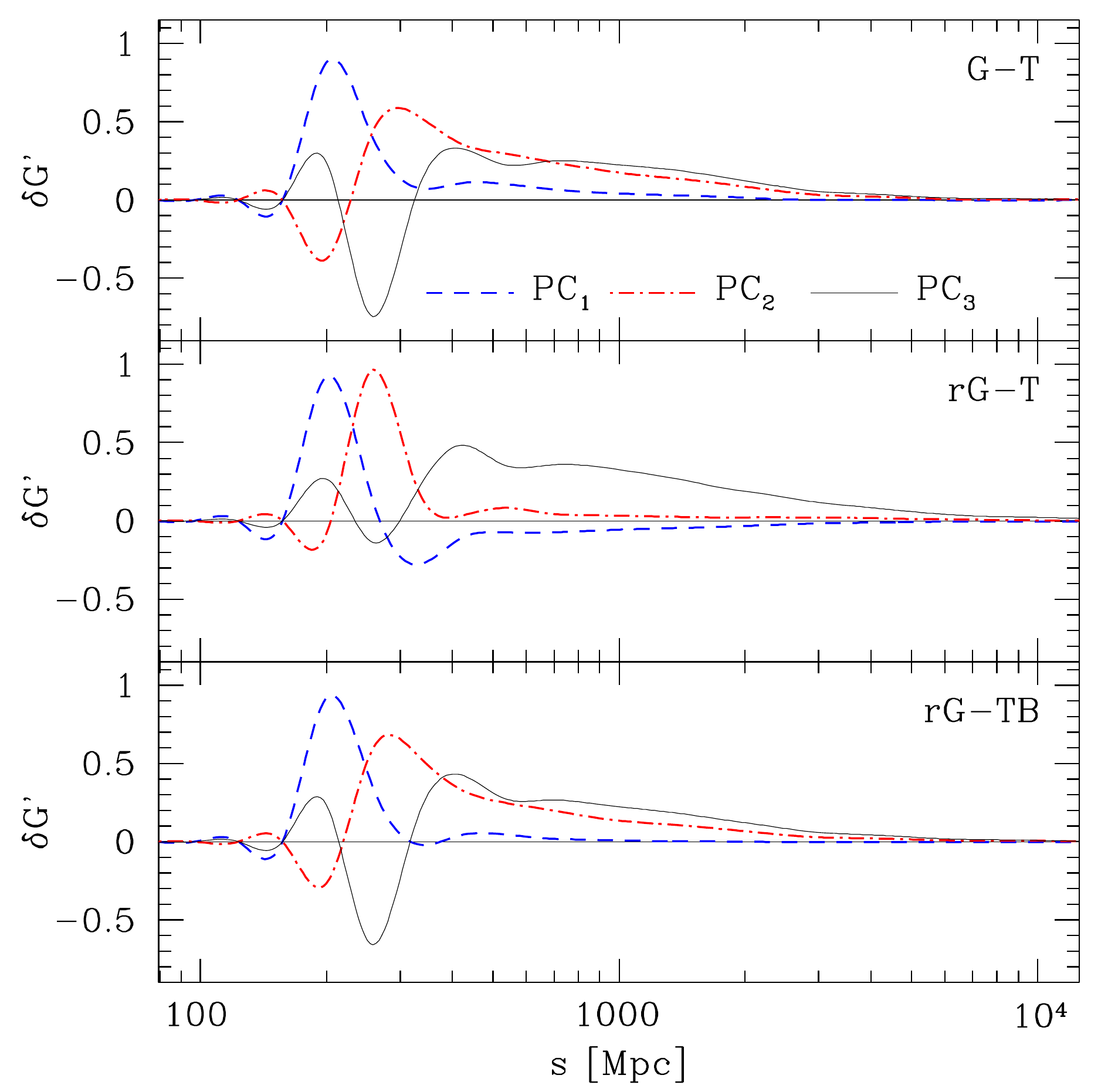, width=3.3in}
\caption{First 3 PC eigenvectors constructed separately for the different model-dataset
combinations.   Although they differ in detail in each combination, the first component mainly determines how rapidly
deviations begin after $s=200$ \,Mpc and the third one carries substantial coherent deviations at
$s>400$\,Mpc.  The second component affects the intermediate regime, and carries different
contributions for
$s>400$ in the different cases.}
\label{fig:pcs}	
\end{figure}

\begin{figure}[h]  
\psfig{file=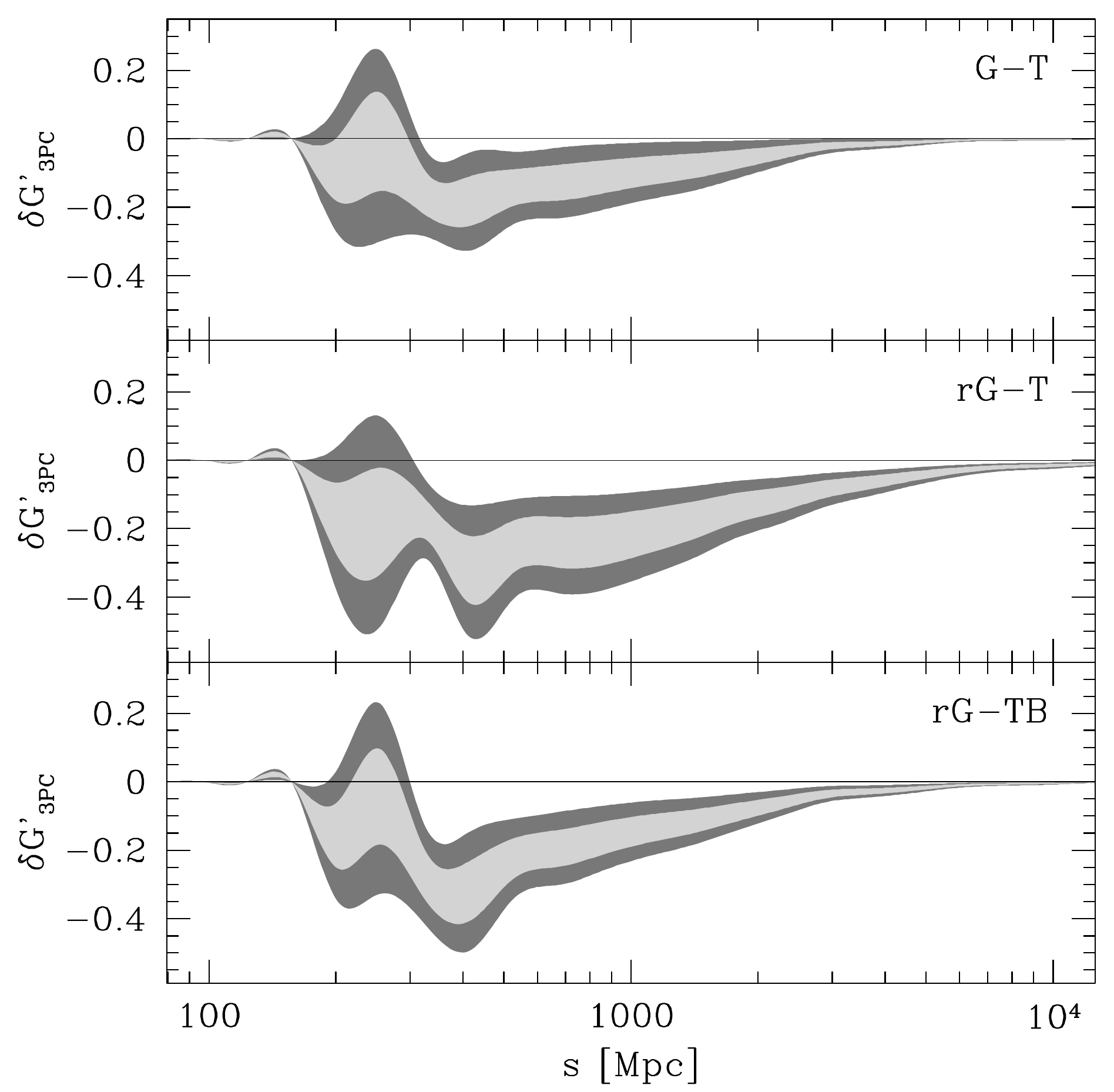, width=3.3in}
\caption{3 PC filtered curvature source $G'_{\rm 3PC}$ (see Eq.~\ref{eqn:PCfilter}).  Favored deviations correspond to a negative source at $s>400$ Mpc whose signficance, depth and extent to
smaller scales increases
for model cases that allow for tensors.
}.
\label{fig:Gprime}	
\end{figure}

  Given the possibility of confirmation by  upcoming polarization measurements,
it is interesting to explore in more detail what constraints on inflationary models these
features 
imply.  
Since the $p_i$ constraints on $\delta G'$ are too noisy to visualize the small but statistically
significant constraints directly, we transform them to the PC basis $m_a$ as 
described in the previous section. 
In Fig.~\ref{fig:pc} we show each of the 20 statistically independent
$m_a$ measurements.  Only the first 3 PCs show measurements that
 that deviate from $m_a=0$ at the 95\%
CL or more.   In terms of the standard errors in Tab.~\ref{table:mc_parameters}, $m_a=0$
is a $2.2 \sigma$  deviation in both $m_1$ and $m_2$.  
Fig.~\ref{fig:PCpost} confirms that $m_a=0$ indeed lies in the tails of the posterior probability
distribution in both.  
 Given the 20 parameter model,
this indicates a preference for a deviation in $\delta G'$ that is significant but not overwhelmingly so.  Given that the corresponding $E$ polarization features can be
twice as large, we can infer that if the $m_a$ parameters remain at their central values,
polarization measurements can provide a convincing detection of the deviation.

These first 3 PCs represent coherent deviations in the source function on scales $s \gtrsim
300$ Mpc with differences mainly reflecting the location and how sharply the deviations rise around that
scale (see Fig.~\ref{fig:pcs}).    
Since the PCs are constructed for each model-dataset independently, a particular $m_a$
does not have a fixed meaning.   It is therefore useful to sum the first three components
together to 
form a 3PC filtered reconstruction of $\delta G'$ from Eq.~(\ref{eqn:PCfilter}) 
shown in Fig.~\ref{fig:Gprime}.   
Of course, more rapid deviations or deviations at $s \gg 10^3$ Mpc 
are allowed by the higher
PCs but they are not significantly constrained by the data.    In fact these models are mainly limited by the prior on $I_1$ in Eq.~(\ref{eqn:I1criterion}) and the sampling rate (see
e.g.~\cite{Adshead:2011jq,Miranda:2013wxa} for allowed models with finer structure).
The data instead favor
a relatively sharp suppression of $\delta G'$ beginning at $s \sim 300-400$ Mpc that is
coherent thereafter.  
  A feature in $\delta G'$ in canonical
$c_s=1$ models  implies a corresponding feature in the inflation potential through 
Eq.~(\ref{eqn:potential}).

\begin{figure}[h]  
\psfig{file=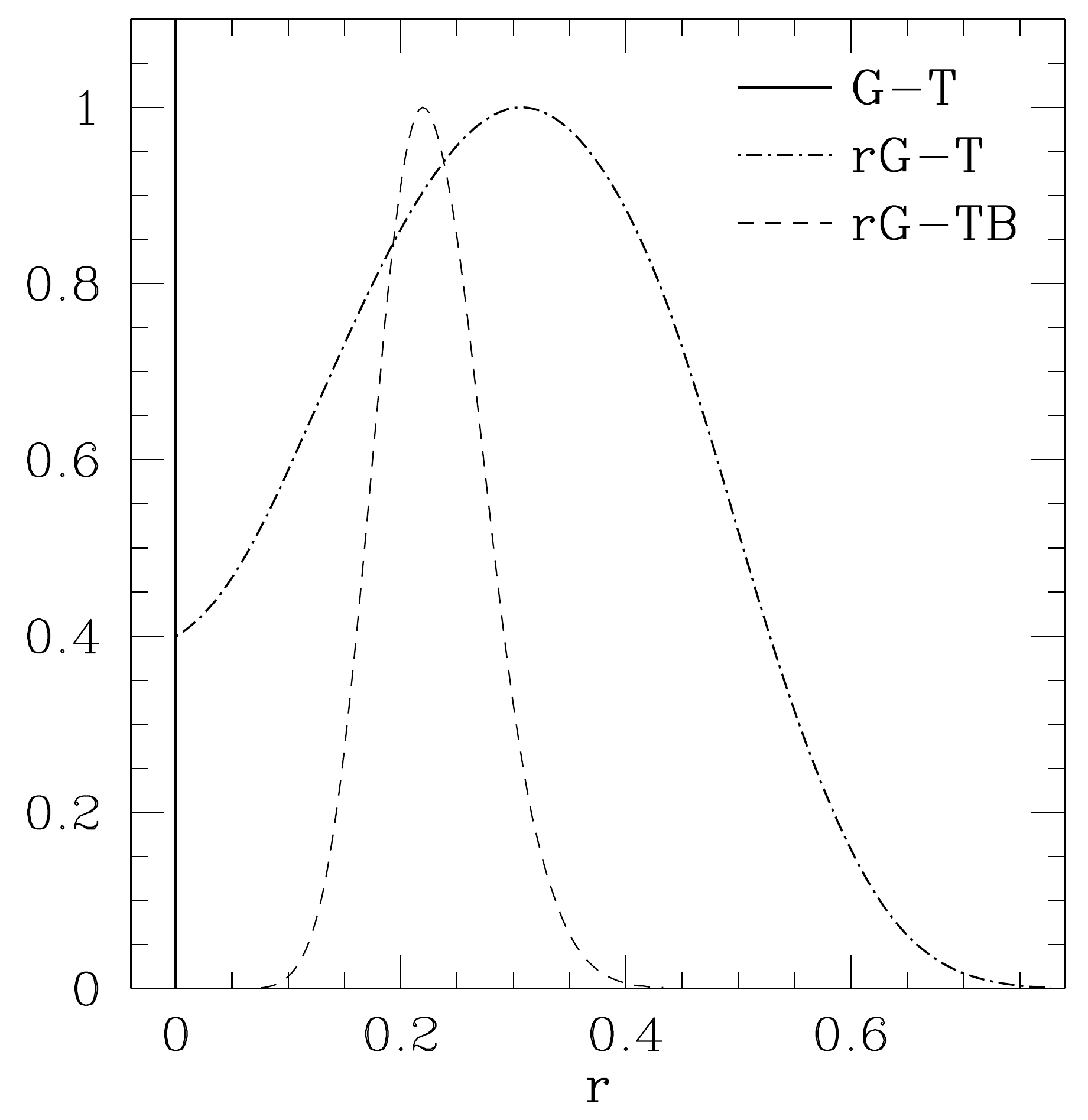, width=2.25in}
\caption{\footnotesize Posterior probability distribution of the tensor-to-scalar ratio $r$ for the various model-dataset combinations.    
In the $G$-$T$ case it is fixed at $r=0$.   In the case that tensors are constrained only by the $T$ dataset, much larger $r$ is allowed
in the $G\Lambda$CDM model compared with the $\Lambda$CDM given the ability to reduce large-scale power in the curvature
spectrum.   The BICEP2 data with no dust contamination favors $r \approx 0.2$, shown here as the $TB$ dataset, and accounting
for contamination still sets a stronger upper limit on $r$. 
}
\label{fig:r}	
\end{figure} 

\subsection{Curvature and Tensors}
\label{sec:tensors}

The preference for features in the curvature source only 
get more significant if the tensor-to-scalar ratio $r$  is allowed
 to vary as in the $rG\Lambda$CDM model.     
 We first consider implications from the temperature-based
 $T$ dataset and call this the $rG$-$T$ analysis. 
 In this dataset, the tensor amplitude is constrained by the shape of
 the temperature power spectrum due to its tensor contributions  above the horizon at
 recombination.    To achieve the same temperature power spectrum, the curvature contributions must be further suppressed and
 hence there is a near degeneracy between $\delta G'$ and $r$.

Thus instead of the upper limit of $r< 0.11$  (95\% CL) 
 at $k=0.002$ Mpc$^{-1}$ \cite{Ade2013}, the constraints on $r$ weaken substantially as shown in
 Fig.~\ref{fig:r}, allowing and even mildly preferring values of $r>0.2$.  
 These large values are mildly preferred because of the excess of power
 in the temperature spectrum around $\ell \sim 40$ (see Fig.~\ref{fig:temp}) which can be explained
 by  a large tensor contribution.   Of course, such an explanation would require an even larger
 suppression of the curvature spectrum on larger scales to produce the same temperature power spectrum.
 
 These qualitative expectations are borne out in the curvature power spectrum constraints in Fig.~\ref{fig:curvature}.
The suppression in the curvature power spectrum begins at 
 $k \gtrsim 0.002$ Mpc$^{-1}$ making the feature there appear less like a glitch
 and more like part of a coherent, but larger and more significant, suppression of long-wavelength power.   Note that the combination of the curvature and tensor sources leads
  to the very similar temperature power spectra shown in Fig.~\ref{fig:temp}.
  
  Interestingly, the prediction for $E$-mode polarization power spectrum  differs qualitatively from the $G$-$T$ case (see Fig.~\ref{fig:pol}).
  Tensors also contribute $E$-modes with a larger $E$ to $T$ ratio than scalars
  due to projection
  effects
  (see e.g.~\cite{Hu:1997hp}, Eq.~25).   Thus instead of a deficit in power there is a  preference
  for an increment in power in the $\ell \sim 20-70$ regime that is allowed to reach in
  excess of $20\%$ in contrast to the $-40\%$ without tensors.  
  Thus, $E$ polarization power spectrum can provide a sharp test of models with $r>0.2$.

 In terms of the principal components, $m_2$ and $m_3=0$ 
 are disfavored at $1.7 \sigma$ and $3.2 \sigma$ respectively in Tab.~\ref{table:mc_parameters} and in Fig.~\ref{fig:pc}, and lie in the tails of the posterior distributions of
 Fig.~\ref{fig:PCpost}.   The first component $m_1$
 no longer shows a significant deviation.   Although the detailed shape of the PCs 
 vary depending on the model-dataset combination (see Fig.~\ref{fig:pcs}), the first component
 is still associated with a rapid change in deviations around $s=200$\,Mpc.   With the addition of
 tensors, the change in the curvature source is more gradual.   
  This can be seen in the 3 PC filtered
 reconstruction of Fig.~\ref{fig:Gprime}, where the main difference is a gradual increase in amplitude to larger $s$
and a broadening of the allowed range.
 
 In fact, the $T$ dataset allows such large values of $r$ that interpreting the BICEP2 $B$-mode
 detection as an upper limit restricts the range of deviations and $E$-polarization predictions.
 In the $rG$-$TB$ analysis, we assume that there is no dust contamination to the measurement and
 hence obtain conservative maximal values for $r$.
 Even under this assumption, the addition of the BICEP2 measurement 
 eliminates models with $r \gtrsim 0.4$ at high confidence
 (see Fig.~\ref{fig:r}).    In the curvature power spectrum, this makes the predictions intermediate
 between the $G$-$T$ and $rG$-$T$ analyses, in particular for $k$ slightly larger than the $k \sim 0.002$ Mpc$^{-1}$ glitch.
Likewise, the $E$-polarization predictions are intermediate as well.  Instead of a deficit or increment
in predicted power, there is little net preference for either.   Note however that in the
$\ell \sim 20-40$ regime there still is a shallower relative dip of $\sim 10-20\%$ which can
still be used to confirm an inflationary feature with precision measurements
 (see also \cite{Miranda:2014wga} for model examples).
 For the PCs of the $rG$-$TB$ analysis the first three components disfavor $m_a=0$ at the $2.2\sigma$, $3.3\sigma$, $2.3\sigma$ levels (see Tab.~\ref{table:mc_parameters} and Figs.~\ref{fig:pc}-\ref{fig:PCpost}).

 Of course, this larger formal significance of should not be interpreted as enhanced evidence for features given
 the uncertain level of contamination by dust.    
   Accounting for some fractional contamination by dust would further interpolate between the $G$-$T$ and $rG$-$TB$ results.   In fact, if a polarization dip at $\ell \sim 20-30$ is detected, its depth relative to the temperature one can be used
 to constrain $r$ further independently of the $B$-modes.

\section{Discussion}
\label{sec:discussion}

We have provided a model-independent analysis of large-scale inflationary features in
the CMB temperature power spectrum allowing for the possibility of non-negligible
tensor contributions.   Unlike similar treatments for the curvature
power spectrum, we directly parameterize the inflationary source of curvature fluctuations
or, correspondingly, features in the potential for canonical single field inflationary models.  
This prevents the problem of fitting the data to unphysical forms for the curvature
power spectrum.   Our parameterization is instead limited by the chosen
1/4 of an efold sampling of temporal features during inflation and the restriction 
to observable scales larger than 200 Mpc. 

When analyzed in terms of the 20 principal components of the curvature source function, 
the temperature anomalies imply deviations from scale-free power law conditions
in 2-3 parameters with 2-3$\sigma$ deviations in each, with larger values reflecting
cases where tensors are allowed.   These deviations correspond to   
a suppression of power at low multipoles beginning with a glitch 
at multipoles $\ell \sim 20-40$.

If tensors are absent, the corresponding $E$-mode polarization features follow
a similar pattern but are predicted to be up to 
twice as large.   They offer the opportunity to double the significance of inflationary 
features or render the temperature anomalies as statistical flukes or inconsistent with
single field inflationary models.
If tensors are allowed, then there is a degeneracy in the temperature power spectrum
between a reduction in curvature fluctuations at large scales and an increase
in the tensor-to-scalar ratio that allows $r>0.2$.   Tensors also change the interpretation
of the glitch by making it more consistent with a monotonic suppression of large-scale curvature power.

 This degeneracy is broken not only
by the $B$-mode polarization of tensors but also by their $E$-mode polarization.   We have
shown that the general signature
of such a large tensor-to-scalar ratio is to predict an increase in the $E$-mode polarization power spectrum
where the deficit in the $T$ power spectrum exists.  
 While the BICEP2 data can 
already be inferred to place an upper bound on $r$ of this order, a precision measurement
of $E$-mode polarization at multipoles from $\ell\sim 20-60$ can potentially provide
an independent constraint that is less subject to dust foreground uncertainties.
These predictions will soon be tested with the release of the $E$-mode polarization
spectrum from the Planck collaboration.

\acknowledgements
 WH and VM  were  supported
 by U.S.~Dept.\ of Energy
 contract DE-FG02-13ER41958
 and the
 Kavli Institute for Cosmological Physics at the University of
 Chicago through grants NSF PHY-0114422 and NSF PHY-0551142.  
 CD was supported by NASA through Hubble Fellowship grant HST-HF2-51340.001 awarded by the Space Telescope Science Institute, which is operated by the Association of Universities for Research in Astronomy, Inc., for NASA, under contract NAS 5-26555. 
This work made use of computing resources and support provided by the Research Computing Center at the University of Chicago.

\vfill 
\break
\bibliography{dG}

\begin{thebibliography}{59}%
\makeatletter
\providecommand \@ifxundefined [1]{%
 \@ifx{#1\undefined}
}%
\providecommand \@ifnum [1]{%
 \ifnum #1\expandafter \@firstoftwo
 \else \expandafter \@secondoftwo
 \fi
}%
\providecommand \@ifx [1]{%
 \ifx #1\expandafter \@firstoftwo
 \else \expandafter \@secondoftwo
 \fi
}%
\providecommand \natexlab [1]{#1}%
\providecommand \enquote  [1]{``#1''}%
\providecommand \bibnamefont  [1]{#1}%
\providecommand \bibfnamefont [1]{#1}%
\providecommand \citenamefont [1]{#1}%
\providecommand \href@noop [0]{\@secondoftwo}%
\providecommand \href [0]{\begingroup \@sanitize@url \@href}%
\providecommand \@href[1]{\@@startlink{#1}\@@href}%
\providecommand \@@href[1]{\endgroup#1\@@endlink}%
\providecommand \@sanitize@url [0]{\catcode `\\12\catcode `\$12\catcode
  `\&12\catcode `\#12\catcode `\^12\catcode `\_12\catcode `\%12\relax}%
\providecommand \@@startlink[1]{}%
\providecommand \@@endlink[0]{}%
\providecommand \url  [0]{\begingroup\@sanitize@url \@url }%
\providecommand \@url [1]{\endgroup\@href {#1}{\urlprefix }}%
\providecommand \urlprefix  [0]{URL }%
\providecommand \Eprint [0]{\href }%
\providecommand \doibase [0]{http://dx.doi.org/}%
\providecommand \selectlanguage [0]{\@gobble}%
\providecommand \bibinfo  [0]{\@secondoftwo}%
\providecommand \bibfield  [0]{\@secondoftwo}%
\providecommand \translation [1]{[#1]}%
\providecommand \BibitemOpen [0]{}%
\providecommand \bibitemStop [0]{}%
\providecommand \bibitemNoStop [0]{.\EOS\space}%
\providecommand \EOS [0]{\spacefactor3000\relax}%
\providecommand \BibitemShut  [1]{\csname bibitem#1\endcsname}%
\let\auto@bib@innerbib\@empty
\bibitem [{\citenamefont {Hinshaw}\ \emph {et~al.}(2003)\citenamefont {Hinshaw}
  \emph {et~al.}}]{Hinshaw:2003ex}%
  \BibitemOpen
  \bibfield  {author} {\bibinfo {author} {\bibfnamefont {G.}~\bibnamefont
  {Hinshaw}} \emph {et~al.} (\bibinfo {collaboration} {WMAP Collaboration}),\
  }\href {\doibase 10.1086/377225} {\bibfield  {journal} {\bibinfo  {journal}
  {Astrophys.J.Suppl.}\ }\textbf {\bibinfo {volume} {148}},\ \bibinfo {pages}
  {135} (\bibinfo {year} {2003})},\ \Eprint
  {http://arxiv.org/abs/astro-ph/0302217} {arXiv:astro-ph/0302217 [astro-ph]}
  \BibitemShut {NoStop}%
\bibitem [{\citenamefont {Peiris}\ \emph {et~al.}(2003)\citenamefont {Peiris}
  \emph {et~al.}}]{Peiris:2003ff}%
  \BibitemOpen
  \bibfield  {author} {\bibinfo {author} {\bibfnamefont {H.}~\bibnamefont
  {Peiris}} \emph {et~al.} (\bibinfo {collaboration} {WMAP Collaboration}),\
  }\href {\doibase 10.1086/377228} {\bibfield  {journal} {\bibinfo  {journal}
  {Astrophys.J.Suppl.}\ }\textbf {\bibinfo {volume} {148}},\ \bibinfo {pages}
  {213} (\bibinfo {year} {2003})},\ \Eprint
  {http://arxiv.org/abs/astro-ph/0302225} {arXiv:astro-ph/0302225 [astro-ph]}
  \BibitemShut {NoStop}%
\bibitem [{\citenamefont {Spergel}\ \emph {et~al.}(2003)\citenamefont {Spergel}
  \emph {et~al.}}]{Spergel:2003cb}%
  \BibitemOpen
  \bibfield  {author} {\bibinfo {author} {\bibfnamefont {D.}~\bibnamefont
  {Spergel}} \emph {et~al.} (\bibinfo {collaboration} {WMAP Collaboration}),\
  }\href {\doibase 10.1086/377226} {\bibfield  {journal} {\bibinfo  {journal}
  {Astrophys.J.Suppl.}\ }\textbf {\bibinfo {volume} {148}},\ \bibinfo {pages}
  {175} (\bibinfo {year} {2003})},\ \Eprint
  {http://arxiv.org/abs/astro-ph/0302209} {arXiv:astro-ph/0302209 [astro-ph]}
  \BibitemShut {NoStop}%
\bibitem [{\citenamefont {Copi}\ \emph {et~al.}(2007)\citenamefont {Copi},
  \citenamefont {Huterer}, \citenamefont {Schwarz},\ and\ \citenamefont
  {Starkman}}]{Copi:2006tu}%
  \BibitemOpen
  \bibfield  {author} {\bibinfo {author} {\bibfnamefont {C.}~\bibnamefont
  {Copi}}, \bibinfo {author} {\bibfnamefont {D.}~\bibnamefont {Huterer}},
  \bibinfo {author} {\bibfnamefont {D.}~\bibnamefont {Schwarz}}, \ and\
  \bibinfo {author} {\bibfnamefont {G.}~\bibnamefont {Starkman}},\ }\href
  {\doibase 10.1103/PhysRevD.75.023507} {\bibfield  {journal} {\bibinfo
  {journal} {Phys.Rev.}\ }\textbf {\bibinfo {volume} {D75}},\ \bibinfo {pages}
  {023507} (\bibinfo {year} {2007})},\ \Eprint
  {http://arxiv.org/abs/astro-ph/0605135} {arXiv:astro-ph/0605135 [astro-ph]}
  \BibitemShut {NoStop}%
\bibitem [{\citenamefont {Ade}\ \emph {et~al.}(2013{\natexlab{a}})\citenamefont
  {Ade} \emph {et~al.}}]{Ade:2013nlj}%
  \BibitemOpen
  \bibfield  {author} {\bibinfo {author} {\bibfnamefont {P.}~\bibnamefont
  {Ade}} \emph {et~al.} (\bibinfo {collaboration} {Planck Collaboration}),\
  }\href@noop {} {\  (\bibinfo {year} {2013}{\natexlab{a}})},\ \Eprint
  {http://arxiv.org/abs/1303.5083} {arXiv:1303.5083 [astro-ph.CO]} \BibitemShut
  {NoStop}%
\bibitem [{\citenamefont {Ade}\ \emph {et~al.}(2014)\citenamefont {Ade} \emph
  {et~al.}}]{Ade:2014xna}%
  \BibitemOpen
  \bibfield  {author} {\bibinfo {author} {\bibfnamefont {P.}~\bibnamefont
  {Ade}} \emph {et~al.} (\bibinfo {collaboration} {BICEP2 Collaboration}),\
  }\href {\doibase 10.1103/PhysRevLett.112.241101} {\bibfield  {journal}
  {\bibinfo  {journal} {Phys.Rev.Lett.}\ }\textbf {\bibinfo {volume} {112}},\
  \bibinfo {pages} {241101} (\bibinfo {year} {2014})},\ \Eprint
  {http://arxiv.org/abs/1403.3985} {arXiv:1403.3985 [astro-ph.CO]} \BibitemShut
  {NoStop}%
\bibitem [{\citenamefont {Adam}\ \emph {et~al.}(2014)\citenamefont {Adam} \emph
  {et~al.}}]{Adam:2014bub}%
  \BibitemOpen
  \bibfield  {author} {\bibinfo {author} {\bibfnamefont {R.}~\bibnamefont
  {Adam}} \emph {et~al.} (\bibinfo {collaboration} {Planck Collaboration}),\
  }\href@noop {} {\  (\bibinfo {year} {2014})},\ \Eprint
  {http://arxiv.org/abs/1409.5738} {arXiv:1409.5738 [astro-ph.CO]} \BibitemShut
  {NoStop}%
\bibitem [{\citenamefont {Ade}\ \emph {et~al.}(2013{\natexlab{b}})\citenamefont
  {Ade} \emph {et~al.}}]{Ade2013}%
  \BibitemOpen
  \bibfield  {author} {\bibinfo {author} {\bibfnamefont {P.}~\bibnamefont
  {Ade}} \emph {et~al.} (\bibinfo {collaboration} {Planck Collaboration}),\
  }\href@noop {} {\  (\bibinfo {year} {2013}{\natexlab{b}})},\ \Eprint
  {http://arxiv.org/abs/1303.5082} {arXiv:1303.5082 [astro-ph.CO]} \BibitemShut
  {NoStop}%
\bibitem [{\citenamefont {Hannestad}(2001)}]{Hannestad:2000pm}%
  \BibitemOpen
  \bibfield  {author} {\bibinfo {author} {\bibfnamefont {S.}~\bibnamefont
  {Hannestad}},\ }\href {\doibase 10.1103/PhysRevD.63.043009} {\bibfield
  {journal} {\bibinfo  {journal} {Phys.Rev.}\ }\textbf {\bibinfo {volume}
  {D63}},\ \bibinfo {pages} {043009} (\bibinfo {year} {2001})},\ \Eprint
  {http://arxiv.org/abs/astro-ph/0009296} {arXiv:astro-ph/0009296 [astro-ph]}
  \BibitemShut {NoStop}%
\bibitem [{\citenamefont {Hu}\ and\ \citenamefont {Okamoto}(2004)}]{Hu:2003vp}%
  \BibitemOpen
  \bibfield  {author} {\bibinfo {author} {\bibfnamefont {W.}~\bibnamefont
  {Hu}}\ and\ \bibinfo {author} {\bibfnamefont {T.}~\bibnamefont {Okamoto}},\
  }\href {\doibase 10.1103/PhysRevD.69.043004} {\bibfield  {journal} {\bibinfo
  {journal} {Phys. Rev.}\ }\textbf {\bibinfo {volume} {D69}},\ \bibinfo {pages}
  {043004} (\bibinfo {year} {2004})},\ \Eprint
  {http://arxiv.org/abs/astro-ph/0308049} {arXiv:astro-ph/0308049} \BibitemShut
  {NoStop}%
\bibitem [{\citenamefont {Tegmark}\ and\ \citenamefont
  {Zaldarriaga}(2002)}]{Tegmark:2002cy}%
  \BibitemOpen
  \bibfield  {author} {\bibinfo {author} {\bibfnamefont {M.}~\bibnamefont
  {Tegmark}}\ and\ \bibinfo {author} {\bibfnamefont {M.}~\bibnamefont
  {Zaldarriaga}},\ }\href {\doibase 10.1103/PhysRevD.66.103508} {\bibfield
  {journal} {\bibinfo  {journal} {Phys.Rev.}\ }\textbf {\bibinfo {volume}
  {D66}},\ \bibinfo {pages} {103508} (\bibinfo {year} {2002})},\ \Eprint
  {http://arxiv.org/abs/astro-ph/0207047} {arXiv:astro-ph/0207047 [astro-ph]}
  \BibitemShut {NoStop}%
\bibitem [{\citenamefont {Hannestad}(2004)}]{Hannestad:2003zs}%
  \BibitemOpen
  \bibfield  {author} {\bibinfo {author} {\bibfnamefont {S.}~\bibnamefont
  {Hannestad}},\ }\href {\doibase 10.1088/1475-7516/2004/04/002} {\bibfield
  {journal} {\bibinfo  {journal} {JCAP}\ }\textbf {\bibinfo {volume} {0404}},\
  \bibinfo {pages} {002} (\bibinfo {year} {2004})},\ \Eprint
  {http://arxiv.org/abs/astro-ph/0311491} {arXiv:astro-ph/0311491 [astro-ph]}
  \BibitemShut {NoStop}%
\bibitem [{\citenamefont {Bridle}\ \emph {et~al.}(2003)\citenamefont {Bridle},
  \citenamefont {Lewis}, \citenamefont {Weller},\ and\ \citenamefont
  {Efstathiou}}]{Bridle:2003sa}%
  \BibitemOpen
  \bibfield  {author} {\bibinfo {author} {\bibfnamefont {S.~L.}\ \bibnamefont
  {Bridle}}, \bibinfo {author} {\bibfnamefont {A.~M.}\ \bibnamefont {Lewis}},
  \bibinfo {author} {\bibfnamefont {J.}~\bibnamefont {Weller}}, \ and\ \bibinfo
  {author} {\bibfnamefont {G.}~\bibnamefont {Efstathiou}},\ }\href@noop {}
  {\bibfield  {journal} {\bibinfo  {journal} {Mon. Not. Roy. Astron. Soc.}\
  }\textbf {\bibinfo {volume} {342}},\ \bibinfo {pages} {L72} (\bibinfo {year}
  {2003})},\ \Eprint {http://arxiv.org/abs/astro-ph/0302306}
  {arXiv:astro-ph/0302306} \BibitemShut {NoStop}%
\bibitem [{\citenamefont {Mukherjee}\ and\ \citenamefont
  {Wang}(2003)}]{Mukherjee:2003ag}%
  \BibitemOpen
  \bibfield  {author} {\bibinfo {author} {\bibfnamefont {P.}~\bibnamefont
  {Mukherjee}}\ and\ \bibinfo {author} {\bibfnamefont {Y.}~\bibnamefont
  {Wang}},\ }\href {\doibase 10.1086/379161} {\bibfield  {journal} {\bibinfo
  {journal} {Astrophys.J.}\ }\textbf {\bibinfo {volume} {599}},\ \bibinfo
  {pages} {1} (\bibinfo {year} {2003})},\ \Eprint
  {http://arxiv.org/abs/astro-ph/0303211} {arXiv:astro-ph/0303211 [astro-ph]}
  \BibitemShut {NoStop}%
\bibitem [{\citenamefont {Leach}(2006)}]{Leach:2005av}%
  \BibitemOpen
  \bibfield  {author} {\bibinfo {author} {\bibfnamefont {S.~M.}\ \bibnamefont
  {Leach}},\ }\href {\doibase 10.1111/j.1365-2966.2006.10842.x} {\bibfield
  {journal} {\bibinfo  {journal} {Mon. Not. Roy. Astron. Soc.}\ }\textbf
  {\bibinfo {volume} {372}},\ \bibinfo {pages} {646} (\bibinfo {year}
  {2006})},\ \Eprint {http://arxiv.org/abs/astro-ph/0506390}
  {arXiv:astro-ph/0506390} \BibitemShut {NoStop}%
\bibitem [{\citenamefont {Peiris}\ and\ \citenamefont
  {Verde}(2010)}]{PhysRevD.81.021302}%
  \BibitemOpen
  \bibfield  {author} {\bibinfo {author} {\bibfnamefont {H.~V.}\ \bibnamefont
  {Peiris}}\ and\ \bibinfo {author} {\bibfnamefont {L.}~\bibnamefont {Verde}},\
  }\href {\doibase 10.1103/PhysRevD.81.021302} {\bibfield  {journal} {\bibinfo
  {journal} {Phys. Rev. D}\ }\textbf {\bibinfo {volume} {81}},\ \bibinfo
  {pages} {021302} (\bibinfo {year} {2010})}\BibitemShut {NoStop}%
\bibitem [{\citenamefont {Hlozek}\ \emph {et~al.}(2012)\citenamefont {Hlozek},
  \citenamefont {Dunkley}, \citenamefont {Addison}, \citenamefont {Appel},
  \citenamefont {Bond} \emph {et~al.}}]{Hlozek:2011pc}%
  \BibitemOpen
  \bibfield  {author} {\bibinfo {author} {\bibfnamefont {R.}~\bibnamefont
  {Hlozek}}, \bibinfo {author} {\bibfnamefont {J.}~\bibnamefont {Dunkley}},
  \bibinfo {author} {\bibfnamefont {G.}~\bibnamefont {Addison}}, \bibinfo
  {author} {\bibfnamefont {J.~W.}\ \bibnamefont {Appel}}, \bibinfo {author}
  {\bibfnamefont {J.~R.}\ \bibnamefont {Bond}},  \emph {et~al.},\ }\href
  {\doibase 10.1088/0004-637X/749/1/90} {\bibfield  {journal} {\bibinfo
  {journal} {Astrophys.J.}\ }\textbf {\bibinfo {volume} {749}},\ \bibinfo
  {pages} {90} (\bibinfo {year} {2012})},\ \Eprint
  {http://arxiv.org/abs/1105.4887} {arXiv:1105.4887 [astro-ph.CO]} \BibitemShut
  {NoStop}%
\bibitem [{\citenamefont {Gauthier}\ and\ \citenamefont
  {Bucher}(2012)}]{Gauthier:2012aq}%
  \BibitemOpen
  \bibfield  {author} {\bibinfo {author} {\bibfnamefont {C.}~\bibnamefont
  {Gauthier}}\ and\ \bibinfo {author} {\bibfnamefont {M.}~\bibnamefont
  {Bucher}},\ }\href {\doibase 10.1088/1475-7516/2012/10/050} {\bibfield
  {journal} {\bibinfo  {journal} {JCAP}\ }\textbf {\bibinfo {volume} {1210}},\
  \bibinfo {pages} {050} (\bibinfo {year} {2012})},\ \Eprint
  {http://arxiv.org/abs/1209.2147} {arXiv:1209.2147 [astro-ph.CO]} \BibitemShut
  {NoStop}%
\bibitem [{\citenamefont {Vazquez}\ \emph {et~al.}(2012)\citenamefont
  {Vazquez}, \citenamefont {Bridges}, \citenamefont {Hobson},\ and\
  \citenamefont {Lasenby}}]{Vazquez:2012ux}%
  \BibitemOpen
  \bibfield  {author} {\bibinfo {author} {\bibfnamefont {J.~A.}\ \bibnamefont
  {Vazquez}}, \bibinfo {author} {\bibfnamefont {M.}~\bibnamefont {Bridges}},
  \bibinfo {author} {\bibfnamefont {M.}~\bibnamefont {Hobson}}, \ and\ \bibinfo
  {author} {\bibfnamefont {A.}~\bibnamefont {Lasenby}},\ }\href {\doibase
  10.1088/1475-7516/2012/06/006} {\bibfield  {journal} {\bibinfo  {journal}
  {JCAP}\ }\textbf {\bibinfo {volume} {1206}},\ \bibinfo {pages} {006}
  (\bibinfo {year} {2012})},\ \Eprint {http://arxiv.org/abs/1203.1252}
  {arXiv:1203.1252 [astro-ph.CO]} \BibitemShut {NoStop}%
\bibitem [{\citenamefont {Hunt}\ and\ \citenamefont
  {Sarkar}(2014)}]{Hunt:2013bha}%
  \BibitemOpen
  \bibfield  {author} {\bibinfo {author} {\bibfnamefont {P.}~\bibnamefont
  {Hunt}}\ and\ \bibinfo {author} {\bibfnamefont {S.}~\bibnamefont {Sarkar}},\
  }\href {\doibase 10.1088/1475-7516/2014/01/025} {\bibfield  {journal}
  {\bibinfo  {journal} {JCAP}\ }\textbf {\bibinfo {volume} {1401}},\ \bibinfo
  {pages} {025} (\bibinfo {year} {2014})},\ \Eprint
  {http://arxiv.org/abs/1308.2317} {arXiv:1308.2317 [astro-ph.CO]} \BibitemShut
  {NoStop}%
\bibitem [{\citenamefont {Aslanyan}\ \emph {et~al.}(2014)\citenamefont
  {Aslanyan}, \citenamefont {Price}, \citenamefont {Abazajian},\ and\
  \citenamefont {Easther}}]{Aslanyan:2014mqa}%
  \BibitemOpen
  \bibfield  {author} {\bibinfo {author} {\bibfnamefont {G.}~\bibnamefont
  {Aslanyan}}, \bibinfo {author} {\bibfnamefont {L.~C.}\ \bibnamefont {Price}},
  \bibinfo {author} {\bibfnamefont {K.~N.}\ \bibnamefont {Abazajian}}, \ and\
  \bibinfo {author} {\bibfnamefont {R.}~\bibnamefont {Easther}},\ }\href
  {\doibase 10.1088/1475-7516/2014/08/052} {\bibfield  {journal} {\bibinfo
  {journal} {JCAP}\ }\textbf {\bibinfo {volume} {1408}},\ \bibinfo {pages}
  {052} (\bibinfo {year} {2014})},\ \Eprint {http://arxiv.org/abs/1403.5849}
  {arXiv:1403.5849 [astro-ph.CO]} \BibitemShut {NoStop}%
\bibitem [{\citenamefont {Hazra}\ \emph
  {et~al.}(2014{\natexlab{a}})\citenamefont {Hazra}, \citenamefont
  {Shafieloo},\ and\ \citenamefont {Souradeep}}]{Hazra:2014jwa}%
  \BibitemOpen
  \bibfield  {author} {\bibinfo {author} {\bibfnamefont {D.~K.}\ \bibnamefont
  {Hazra}}, \bibinfo {author} {\bibfnamefont {A.}~\bibnamefont {Shafieloo}}, \
  and\ \bibinfo {author} {\bibfnamefont {T.}~\bibnamefont {Souradeep}},\
  }\href@noop {} {\  (\bibinfo {year} {2014}{\natexlab{a}})},\ \Eprint
  {http://arxiv.org/abs/1406.4827} {arXiv:1406.4827 [astro-ph.CO]} \BibitemShut
  {NoStop}%
\bibitem [{\citenamefont {Contaldi}\ \emph {et~al.}(2003)\citenamefont
  {Contaldi}, \citenamefont {Peloso}, \citenamefont {Kofman},\ and\
  \citenamefont {Linde}}]{Contaldi:2003zv}%
  \BibitemOpen
  \bibfield  {author} {\bibinfo {author} {\bibfnamefont {C.~R.}\ \bibnamefont
  {Contaldi}}, \bibinfo {author} {\bibfnamefont {M.}~\bibnamefont {Peloso}},
  \bibinfo {author} {\bibfnamefont {L.}~\bibnamefont {Kofman}}, \ and\ \bibinfo
  {author} {\bibfnamefont {A.~D.}\ \bibnamefont {Linde}},\ }\href {\doibase
  10.1088/1475-7516/2003/07/002} {\bibfield  {journal} {\bibinfo  {journal}
  {JCAP}\ }\textbf {\bibinfo {volume} {0307}},\ \bibinfo {pages} {002}
  (\bibinfo {year} {2003})},\ \Eprint {http://arxiv.org/abs/astro-ph/0303636}
  {arXiv:astro-ph/0303636 [astro-ph]} \BibitemShut {NoStop}%
\bibitem [{\citenamefont {Martin}\ and\ \citenamefont
  {Ringeval}(2004)}]{Martin:2003sg}%
  \BibitemOpen
  \bibfield  {author} {\bibinfo {author} {\bibfnamefont {J.}~\bibnamefont
  {Martin}}\ and\ \bibinfo {author} {\bibfnamefont {C.}~\bibnamefont
  {Ringeval}},\ }\href {\doibase 10.1103/PhysRevD.69.083515} {\bibfield
  {journal} {\bibinfo  {journal} {Phys.Rev.}\ }\textbf {\bibinfo {volume}
  {D69}},\ \bibinfo {pages} {083515} (\bibinfo {year} {2004})},\ \Eprint
  {http://arxiv.org/abs/astro-ph/0310382} {arXiv:astro-ph/0310382 [astro-ph]}
  \BibitemShut {NoStop}%
\bibitem [{\citenamefont {Freivogel}\ \emph {et~al.}(2006)\citenamefont
  {Freivogel}, \citenamefont {Kleban}, \citenamefont {Rodriguez~Martinez},\
  and\ \citenamefont {Susskind}}]{Freivogel:2005vv}%
  \BibitemOpen
  \bibfield  {author} {\bibinfo {author} {\bibfnamefont {B.}~\bibnamefont
  {Freivogel}}, \bibinfo {author} {\bibfnamefont {M.}~\bibnamefont {Kleban}},
  \bibinfo {author} {\bibfnamefont {M.}~\bibnamefont {Rodriguez~Martinez}}, \
  and\ \bibinfo {author} {\bibfnamefont {L.}~\bibnamefont {Susskind}},\ }\href
  {\doibase 10.1088/1126-6708/2006/03/039} {\bibfield  {journal} {\bibinfo
  {journal} {JHEP}\ }\textbf {\bibinfo {volume} {0603}},\ \bibinfo {pages}
  {039} (\bibinfo {year} {2006})},\ \Eprint
  {http://arxiv.org/abs/hep-th/0505232} {arXiv:hep-th/0505232 [hep-th]}
  \BibitemShut {NoStop}%
\bibitem [{\citenamefont {Covi}\ \emph {et~al.}(2006)\citenamefont {Covi},
  \citenamefont {Hamann}, \citenamefont {Melchiorri}, \citenamefont {Slosar},\
  and\ \citenamefont {Sorbera}}]{Covi:2006ci}%
  \BibitemOpen
  \bibfield  {author} {\bibinfo {author} {\bibfnamefont {L.}~\bibnamefont
  {Covi}}, \bibinfo {author} {\bibfnamefont {J.}~\bibnamefont {Hamann}},
  \bibinfo {author} {\bibfnamefont {A.}~\bibnamefont {Melchiorri}}, \bibinfo
  {author} {\bibfnamefont {A.}~\bibnamefont {Slosar}}, \ and\ \bibinfo {author}
  {\bibfnamefont {I.}~\bibnamefont {Sorbera}},\ }\href {\doibase
  10.1103/PhysRevD.74.083509} {\bibfield  {journal} {\bibinfo  {journal} {Phys.
  Rev.}\ }\textbf {\bibinfo {volume} {D74}},\ \bibinfo {pages} {083509}
  (\bibinfo {year} {2006})},\ \Eprint {http://arxiv.org/abs/astro-ph/0606452}
  {arXiv:astro-ph/0606452} \BibitemShut {NoStop}%
\bibitem [{\citenamefont {Joy}\ \emph {et~al.}(2008)\citenamefont {Joy},
  \citenamefont {Sahni},\ and\ \citenamefont {Starobinsky}}]{Joy:2007na}%
  \BibitemOpen
  \bibfield  {author} {\bibinfo {author} {\bibfnamefont {M.}~\bibnamefont
  {Joy}}, \bibinfo {author} {\bibfnamefont {V.}~\bibnamefont {Sahni}}, \ and\
  \bibinfo {author} {\bibfnamefont {A.~A.}\ \bibnamefont {Starobinsky}},\
  }\href {\doibase 10.1103/PhysRevD.77.023514} {\bibfield  {journal} {\bibinfo
  {journal} {Phys.Rev.}\ }\textbf {\bibinfo {volume} {D77}},\ \bibinfo {pages}
  {023514} (\bibinfo {year} {2008})},\ \Eprint {http://arxiv.org/abs/0711.1585}
  {arXiv:0711.1585 [astro-ph]} \BibitemShut {NoStop}%
\bibitem [{\citenamefont {Hazra}\ \emph {et~al.}(2010)\citenamefont {Hazra},
  \citenamefont {Aich}, \citenamefont {Jain}, \citenamefont {Sriramkumar},\
  and\ \citenamefont {Souradeep}}]{Hazra:2010ve}%
  \BibitemOpen
  \bibfield  {author} {\bibinfo {author} {\bibfnamefont {D.~K.}\ \bibnamefont
  {Hazra}}, \bibinfo {author} {\bibfnamefont {M.}~\bibnamefont {Aich}},
  \bibinfo {author} {\bibfnamefont {R.~K.}\ \bibnamefont {Jain}}, \bibinfo
  {author} {\bibfnamefont {L.}~\bibnamefont {Sriramkumar}}, \ and\ \bibinfo
  {author} {\bibfnamefont {T.}~\bibnamefont {Souradeep}},\ }\href {\doibase
  10.1088/1475-7516/2010/10/008} {\bibfield  {journal} {\bibinfo  {journal}
  {JCAP}\ }\textbf {\bibinfo {volume} {1010}},\ \bibinfo {pages} {008}
  (\bibinfo {year} {2010})},\ \Eprint {http://arxiv.org/abs/1005.2175}
  {arXiv:1005.2175 [astro-ph.CO]} \BibitemShut {NoStop}%
\bibitem [{\citenamefont {Achucarro}\ \emph {et~al.}(2014)\citenamefont
  {Achucarro}, \citenamefont {Atal}, \citenamefont {Ortiz},\ and\ \citenamefont
  {Torrado}}]{Achucarro:2013cva}%
  \BibitemOpen
  \bibfield  {author} {\bibinfo {author} {\bibfnamefont {A.}~\bibnamefont
  {Achucarro}}, \bibinfo {author} {\bibfnamefont {V.}~\bibnamefont {Atal}},
  \bibinfo {author} {\bibfnamefont {P.}~\bibnamefont {Ortiz}}, \ and\ \bibinfo
  {author} {\bibfnamefont {J.}~\bibnamefont {Torrado}},\ }\href {\doibase
  10.1103/PhysRevD.89.103006} {\bibfield  {journal} {\bibinfo  {journal}
  {Phys.Rev.}\ }\textbf {\bibinfo {volume} {D89}},\ \bibinfo {pages} {103006}
  (\bibinfo {year} {2014})},\ \Eprint {http://arxiv.org/abs/1311.2552}
  {arXiv:1311.2552 [astro-ph.CO]} \BibitemShut {NoStop}%
\bibitem [{\citenamefont {Contaldi}\ \emph {et~al.}(2014)\citenamefont
  {Contaldi}, \citenamefont {Peloso},\ and\ \citenamefont
  {Sorbo}}]{Contaldi:2014zua}%
  \BibitemOpen
  \bibfield  {author} {\bibinfo {author} {\bibfnamefont {C.~R.}\ \bibnamefont
  {Contaldi}}, \bibinfo {author} {\bibfnamefont {M.}~\bibnamefont {Peloso}}, \
  and\ \bibinfo {author} {\bibfnamefont {L.}~\bibnamefont {Sorbo}},\
  }\href@noop {} {\  (\bibinfo {year} {2014})},\ \Eprint
  {http://arxiv.org/abs/1403.4596} {arXiv:1403.4596 [astro-ph.CO]} \BibitemShut
  {NoStop}%
\bibitem [{\citenamefont {Miranda}\ \emph {et~al.}(2014)\citenamefont
  {Miranda}, \citenamefont {Hu},\ and\ \citenamefont
  {Adshead}}]{Miranda:2014wga}%
  \BibitemOpen
  \bibfield  {author} {\bibinfo {author} {\bibfnamefont {V.}~\bibnamefont
  {Miranda}}, \bibinfo {author} {\bibfnamefont {W.}~\bibnamefont {Hu}}, \ and\
  \bibinfo {author} {\bibfnamefont {P.}~\bibnamefont {Adshead}},\ }\href@noop
  {} {\bibfield  {journal} {\bibinfo  {journal} {Phys.Rev.}\ }\textbf {\bibinfo
  {volume} {D89}},\ \bibinfo {pages} {101302} (\bibinfo {year} {2014})},\
  \Eprint {http://arxiv.org/abs/1403.5231} {arXiv:1403.5231 [astro-ph.CO]}
  \BibitemShut {NoStop}%
\bibitem [{\citenamefont {Abazajian}\ \emph {et~al.}(2014)\citenamefont
  {Abazajian}, \citenamefont {Aslanyan}, \citenamefont {Easther},\ and\
  \citenamefont {Price}}]{Abazajian:2014tqa}%
  \BibitemOpen
  \bibfield  {author} {\bibinfo {author} {\bibfnamefont {K.~N.}\ \bibnamefont
  {Abazajian}}, \bibinfo {author} {\bibfnamefont {G.}~\bibnamefont {Aslanyan}},
  \bibinfo {author} {\bibfnamefont {R.}~\bibnamefont {Easther}}, \ and\
  \bibinfo {author} {\bibfnamefont {L.~C.}\ \bibnamefont {Price}},\ }\href
  {\doibase 10.1088/1475-7516/2014/08/053} {\bibfield  {journal} {\bibinfo
  {journal} {JCAP}\ }\textbf {\bibinfo {volume} {1408}},\ \bibinfo {pages}
  {053} (\bibinfo {year} {2014})},\ \Eprint {http://arxiv.org/abs/1403.5922}
  {arXiv:1403.5922 [astro-ph.CO]} \BibitemShut {NoStop}%
\bibitem [{\citenamefont {Hazra}\ \emph
  {et~al.}(2014{\natexlab{b}})\citenamefont {Hazra}, \citenamefont {Shafieloo},
  \citenamefont {Smoot},\ and\ \citenamefont {Starobinsky}}]{Hazra:2014jka}%
  \BibitemOpen
  \bibfield  {author} {\bibinfo {author} {\bibfnamefont {D.~K.}\ \bibnamefont
  {Hazra}}, \bibinfo {author} {\bibfnamefont {A.}~\bibnamefont {Shafieloo}},
  \bibinfo {author} {\bibfnamefont {G.~F.}\ \bibnamefont {Smoot}}, \ and\
  \bibinfo {author} {\bibfnamefont {A.~A.}\ \bibnamefont {Starobinsky}},\
  }\href@noop {} {\  (\bibinfo {year} {2014}{\natexlab{b}})},\ \Eprint
  {http://arxiv.org/abs/1404.0360} {arXiv:1404.0360 [astro-ph.CO]} \BibitemShut
  {NoStop}%
\bibitem [{\citenamefont {Bousso}\ \emph {et~al.}(2014)\citenamefont {Bousso},
  \citenamefont {Harlow},\ and\ \citenamefont {Senatore}}]{Bousso:2014jca}%
  \BibitemOpen
  \bibfield  {author} {\bibinfo {author} {\bibfnamefont {R.}~\bibnamefont
  {Bousso}}, \bibinfo {author} {\bibfnamefont {D.}~\bibnamefont {Harlow}}, \
  and\ \bibinfo {author} {\bibfnamefont {L.}~\bibnamefont {Senatore}},\
  }\href@noop {} {\  (\bibinfo {year} {2014})},\ \Eprint
  {http://arxiv.org/abs/1404.2278} {arXiv:1404.2278 [astro-ph.CO]} \BibitemShut
  {NoStop}%
\bibitem [{\citenamefont {Mortonson}\ \emph {et~al.}(2009)\citenamefont
  {Mortonson}, \citenamefont {Dvorkin}, \citenamefont {Peiris},\ and\
  \citenamefont {Hu}}]{Mortonson2009}%
  \BibitemOpen
  \bibfield  {author} {\bibinfo {author} {\bibfnamefont {M.~J.}\ \bibnamefont
  {Mortonson}}, \bibinfo {author} {\bibfnamefont {C.}~\bibnamefont {Dvorkin}},
  \bibinfo {author} {\bibfnamefont {H.~V.}\ \bibnamefont {Peiris}}, \ and\
  \bibinfo {author} {\bibfnamefont {W.}~\bibnamefont {Hu}},\ }\href {\doibase
  10.1103/PhysRevD.79.103519} {\bibfield  {journal} {\bibinfo  {journal} {Phys.
  Rev.}\ }\textbf {\bibinfo {volume} {D79}},\ \bibinfo {pages} {103519}
  (\bibinfo {year} {2009})},\ \Eprint {http://arxiv.org/abs/0903.4920}
  {arXiv:0903.4920 [astro-ph.CO]} \BibitemShut {NoStop}%
\bibitem [{\citenamefont {Stewart}(2002{\natexlab{a}})}]{Stewart2002}%
  \BibitemOpen
  \bibfield  {author} {\bibinfo {author} {\bibfnamefont {E.~D.}\ \bibnamefont
  {Stewart}},\ }\href {\doibase 10.1103/PhysRevD.65.103508} {\bibfield
  {journal} {\bibinfo  {journal} {Phys. Rev.}\ }\textbf {\bibinfo {volume}
  {D65}},\ \bibinfo {pages} {103508} (\bibinfo {year} {2002}{\natexlab{a}})},\
  \Eprint {http://arxiv.org/abs/astro-ph/0110322} {arXiv:astro-ph/0110322}
  \BibitemShut {NoStop}%
\bibitem [{\citenamefont {Choe}\ \emph
  {et~al.}(2004{\natexlab{a}})\citenamefont {Choe}, \citenamefont {Gong},\ and\
  \citenamefont {Stewart}}]{Choe2004}%
  \BibitemOpen
  \bibfield  {author} {\bibinfo {author} {\bibfnamefont {J.}~\bibnamefont
  {Choe}}, \bibinfo {author} {\bibfnamefont {J.-O.}\ \bibnamefont {Gong}}, \
  and\ \bibinfo {author} {\bibfnamefont {E.~D.}\ \bibnamefont {Stewart}},\
  }\href {\doibase 10.1088/1475-7516/2004/07/012} {\bibfield  {journal}
  {\bibinfo  {journal} {JCAP}\ }\textbf {\bibinfo {volume} {0407}},\ \bibinfo
  {pages} {012} (\bibinfo {year} {2004}{\natexlab{a}})},\ \Eprint
  {http://arxiv.org/abs/hep-ph/0405155} {arXiv:hep-ph/0405155} \BibitemShut
  {NoStop}%
\bibitem [{\citenamefont {Dvorkin}\ and\ \citenamefont
  {Hu}(2010{\natexlab{a}})}]{Dvorkin2010}%
  \BibitemOpen
  \bibfield  {author} {\bibinfo {author} {\bibfnamefont {C.}~\bibnamefont
  {Dvorkin}}\ and\ \bibinfo {author} {\bibfnamefont {W.}~\bibnamefont {Hu}},\
  }\href {\doibase 10.1103/PhysRevD.81.023518} {\bibfield  {journal} {\bibinfo
  {journal} {Phys. Rev.}\ }\textbf {\bibinfo {volume} {D81}},\ \bibinfo {pages}
  {023518} (\bibinfo {year} {2010}{\natexlab{a}})},\ \Eprint
  {http://arxiv.org/abs/0910.2237} {arXiv:0910.2237 [astro-ph.CO]} \BibitemShut
  {NoStop}%
\bibitem [{\citenamefont {Dvorkin}\ and\ \citenamefont
  {Hu}(2010{\natexlab{b}})}]{DvoHu10a}%
  \BibitemOpen
  \bibfield  {author} {\bibinfo {author} {\bibfnamefont {C.}~\bibnamefont
  {Dvorkin}}\ and\ \bibinfo {author} {\bibfnamefont {W.}~\bibnamefont {Hu}},\
  }\href {\doibase 10.1103/PhysRevD.82.043513} {\bibfield  {journal} {\bibinfo
  {journal} {Phys. Rev.}\ }\textbf {\bibinfo {volume} {D82}},\ \bibinfo {pages}
  {043513} (\bibinfo {year} {2010}{\natexlab{b}})},\ \Eprint
  {http://arxiv.org/abs/1007.0215} {arXiv:1007.0215 [astro-ph.CO]} \BibitemShut
  {NoStop}%
\bibitem [{\citenamefont {Dvorkin}\ and\ \citenamefont
  {Hu}(2011)}]{Dvorkin:2011ui}%
  \BibitemOpen
  \bibfield  {author} {\bibinfo {author} {\bibfnamefont {C.}~\bibnamefont
  {Dvorkin}}\ and\ \bibinfo {author} {\bibfnamefont {W.}~\bibnamefont {Hu}},\
  }\href {\doibase 10.1103/PhysRevD.84.063515} {\bibfield  {journal} {\bibinfo
  {journal} {Phys.Rev.}\ }\textbf {\bibinfo {volume} {D84}},\ \bibinfo {pages}
  {063515} (\bibinfo {year} {2011})},\ \Eprint {http://arxiv.org/abs/1106.4016}
  {arXiv:1106.4016 [astro-ph.CO]} \BibitemShut {NoStop}%
\bibitem [{\citenamefont {Stewart}(2002{\natexlab{b}})}]{Stewart:2001cd}%
  \BibitemOpen
  \bibfield  {author} {\bibinfo {author} {\bibfnamefont {E.~D.}\ \bibnamefont
  {Stewart}},\ }\href {\doibase 10.1103/PhysRevD.65.103508} {\bibfield
  {journal} {\bibinfo  {journal} {Phys.Rev.}\ }\textbf {\bibinfo {volume}
  {D65}},\ \bibinfo {pages} {103508} (\bibinfo {year} {2002}{\natexlab{b}})},\
  \Eprint {http://arxiv.org/abs/astro-ph/0110322} {arXiv:astro-ph/0110322
  [astro-ph]} \BibitemShut {NoStop}%
\bibitem [{\citenamefont {Hu}(2011{\natexlab{a}})}]{Hu:2011vr}%
  \BibitemOpen
  \bibfield  {author} {\bibinfo {author} {\bibfnamefont {W.}~\bibnamefont
  {Hu}},\ }\href {\doibase 10.1103/PhysRevD.84.027303} {\bibfield  {journal}
  {\bibinfo  {journal} {Phys.Rev.}\ }\textbf {\bibinfo {volume} {D84}},\
  \bibinfo {pages} {027303} (\bibinfo {year} {2011}{\natexlab{a}})},\ \Eprint
  {http://arxiv.org/abs/1104.4500} {arXiv:1104.4500 [astro-ph.CO]} \BibitemShut
  {NoStop}%
\bibitem [{\citenamefont {Choe}\ \emph
  {et~al.}(2004{\natexlab{b}})\citenamefont {Choe}, \citenamefont {Gong},\ and\
  \citenamefont {Stewart}}]{Choe:2004zg}%
  \BibitemOpen
  \bibfield  {author} {\bibinfo {author} {\bibfnamefont {J.}~\bibnamefont
  {Choe}}, \bibinfo {author} {\bibfnamefont {J.-O.}\ \bibnamefont {Gong}}, \
  and\ \bibinfo {author} {\bibfnamefont {E.~D.}\ \bibnamefont {Stewart}},\
  }\href {\doibase 10.1088/1475-7516/2004/07/012} {\bibfield  {journal}
  {\bibinfo  {journal} {JCAP}\ }\textbf {\bibinfo {volume} {0407}},\ \bibinfo
  {pages} {012} (\bibinfo {year} {2004}{\natexlab{b}})},\ \Eprint
  {http://arxiv.org/abs/hep-ph/0405155} {arXiv:hep-ph/0405155 [hep-ph]}
  \BibitemShut {NoStop}%
\bibitem [{\citenamefont {Dvorkin}\ and\ \citenamefont
  {Hu}(2010{\natexlab{c}})}]{Dvorkin:2009ne}%
  \BibitemOpen
  \bibfield  {author} {\bibinfo {author} {\bibfnamefont {C.}~\bibnamefont
  {Dvorkin}}\ and\ \bibinfo {author} {\bibfnamefont {W.}~\bibnamefont {Hu}},\
  }\href {\doibase 10.1103/PhysRevD.81.023518} {\bibfield  {journal} {\bibinfo
  {journal} {Phys.Rev.}\ }\textbf {\bibinfo {volume} {D81}},\ \bibinfo {pages}
  {023518} (\bibinfo {year} {2010}{\natexlab{c}})},\ \Eprint
  {http://arxiv.org/abs/0910.2237} {arXiv:0910.2237 [astro-ph.CO]} \BibitemShut
  {NoStop}%
\bibitem [{\citenamefont {Hu}(2011{\natexlab{b}})}]{Hu2011}%
  \BibitemOpen
  \bibfield  {author} {\bibinfo {author} {\bibfnamefont {W.}~\bibnamefont
  {Hu}},\ }\href {\doibase 10.1103/PhysRevD.84.027303} {\bibfield  {journal}
  {\bibinfo  {journal} {Phys.Rev.}\ }\textbf {\bibinfo {volume} {D84}},\
  \bibinfo {pages} {027303} (\bibinfo {year} {2011}{\natexlab{b}})},\ \Eprint
  {http://arxiv.org/abs/1104.4500} {arXiv:1104.4500 [astro-ph.CO]} \BibitemShut
  {NoStop}%
\bibitem [{\citenamefont {Cheung}\ \emph {et~al.}(2008)\citenamefont {Cheung},
  \citenamefont {Creminelli}, \citenamefont {Fitzpatrick}, \citenamefont
  {Kaplan},\ and\ \citenamefont {Senatore}}]{Cheung:2007st}%
  \BibitemOpen
  \bibfield  {author} {\bibinfo {author} {\bibfnamefont {C.}~\bibnamefont
  {Cheung}}, \bibinfo {author} {\bibfnamefont {P.}~\bibnamefont {Creminelli}},
  \bibinfo {author} {\bibfnamefont {A.~L.}\ \bibnamefont {Fitzpatrick}},
  \bibinfo {author} {\bibfnamefont {J.}~\bibnamefont {Kaplan}}, \ and\ \bibinfo
  {author} {\bibfnamefont {L.}~\bibnamefont {Senatore}},\ }\href {\doibase
  10.1088/1126-6708/2008/03/014} {\bibfield  {journal} {\bibinfo  {journal}
  {JHEP}\ }\textbf {\bibinfo {volume} {0803}},\ \bibinfo {pages} {014}
  (\bibinfo {year} {2008})},\ \Eprint {http://arxiv.org/abs/0709.0293}
  {arXiv:0709.0293 [hep-th]} \BibitemShut {NoStop}%
\bibitem [{\citenamefont {Gong}(2004)}]{Gong:2004kd}%
  \BibitemOpen
  \bibfield  {author} {\bibinfo {author} {\bibfnamefont {J.-O.}\ \bibnamefont
  {Gong}},\ }\href {\doibase 10.1088/0264-9381/21/23/016} {\bibfield  {journal}
  {\bibinfo  {journal} {Class.Quant.Grav.}\ }\textbf {\bibinfo {volume} {21}},\
  \bibinfo {pages} {5555} (\bibinfo {year} {2004})},\ \Eprint
  {http://arxiv.org/abs/gr-qc/0408039} {arXiv:gr-qc/0408039 [gr-qc]}
  \BibitemShut {NoStop}%
\bibitem [{\citenamefont {Hu}(2014)}]{Hu:2014hoa}%
  \BibitemOpen
  \bibfield  {author} {\bibinfo {author} {\bibfnamefont {W.}~\bibnamefont
  {Hu}},\ }\href {\doibase 10.1103/PhysRevD.89.123503} {\bibfield  {journal}
  {\bibinfo  {journal} {Phys.Rev.}\ }\textbf {\bibinfo {volume} {D89}},\
  \bibinfo {pages} {123503} (\bibinfo {year} {2014})},\ \Eprint
  {http://arxiv.org/abs/1405.2020} {arXiv:1405.2020 [astro-ph.CO]} \BibitemShut
  {NoStop}%
\bibitem [{\citenamefont {Miranda}\ and\ \citenamefont
  {Hu}(2014)}]{Miranda:2013wxa}%
  \BibitemOpen
  \bibfield  {author} {\bibinfo {author} {\bibfnamefont {V.}~\bibnamefont
  {Miranda}}\ and\ \bibinfo {author} {\bibfnamefont {W.}~\bibnamefont {Hu}},\
  }\href {\doibase 10.1103/PhysRevD.89.083529} {\bibfield  {journal} {\bibinfo
  {journal} {Phys.Rev.}\ }\textbf {\bibinfo {volume} {D89}},\ \bibinfo {pages}
  {083529} (\bibinfo {year} {2014})},\ \Eprint {http://arxiv.org/abs/1312.0946}
  {arXiv:1312.0946 [astro-ph.CO]} \BibitemShut {NoStop}%
\bibitem [{\citenamefont {Lewis}\ and\ \citenamefont
  {Bridle}(2002)}]{Lewis:2002ah}%
  \BibitemOpen
  \bibfield  {author} {\bibinfo {author} {\bibfnamefont {A.}~\bibnamefont
  {Lewis}}\ and\ \bibinfo {author} {\bibfnamefont {S.}~\bibnamefont {Bridle}},\
  }\href@noop {} {\bibfield  {journal} {\bibinfo  {journal} {Phys. Rev.}\
  }\textbf {\bibinfo {volume} {D66}},\ \bibinfo {pages} {103511} (\bibinfo
  {year} {2002})},\ \Eprint {http://arxiv.org/abs/astro-ph/0205436}
  {astro-ph/0205436} \BibitemShut {NoStop}%
\bibitem [{\citenamefont {Ade}\ \emph {et~al.}(2013{\natexlab{c}})\citenamefont
  {Ade} \emph {et~al.}}]{Ade:2013zuv}%
  \BibitemOpen
  \bibfield  {author} {\bibinfo {author} {\bibfnamefont {P.}~\bibnamefont
  {Ade}} \emph {et~al.} (\bibinfo {collaboration} {Planck Collaboration}),\
  }\href@noop {} {\  (\bibinfo {year} {2013}{\natexlab{c}})},\ \Eprint
  {http://arxiv.org/abs/1303.5076} {arXiv:1303.5076 [astro-ph.CO]} \BibitemShut
  {NoStop}%
\bibitem [{\citenamefont {Bennett}\ \emph {et~al.}(2012)\citenamefont
  {Bennett}, \citenamefont {Larson}, \citenamefont {Weiland}, \citenamefont
  {Jarosik}, \citenamefont {Hinshaw} \emph {et~al.}}]{Bennett:2012fp}%
  \BibitemOpen
  \bibfield  {author} {\bibinfo {author} {\bibfnamefont {C.}~\bibnamefont
  {Bennett}}, \bibinfo {author} {\bibfnamefont {D.}~\bibnamefont {Larson}},
  \bibinfo {author} {\bibfnamefont {J.}~\bibnamefont {Weiland}}, \bibinfo
  {author} {\bibfnamefont {N.}~\bibnamefont {Jarosik}}, \bibinfo {author}
  {\bibfnamefont {G.}~\bibnamefont {Hinshaw}},  \emph {et~al.},\ }\href@noop {}
  {\  (\bibinfo {year} {2012})},\ \Eprint {http://arxiv.org/abs/1212.5225}
  {arXiv:1212.5225 [astro-ph.CO]} \BibitemShut {NoStop}%
\bibitem [{Note1()}]{Note1}%
  \BibitemOpen
  \bibinfo {note} {\protect \href
  {http://www.supernova.lbl.gov/Union}{http://www.supernova.lbl.gov/Union}}\BibitemShut
  {NoStop}%
\bibitem [{\citenamefont {Anderson}\ \emph {et~al.}(2013)\citenamefont
  {Anderson}, \citenamefont {Aubourg}, \citenamefont {Bailey}, \citenamefont
  {Bizyaev}, \citenamefont {Blanton} \emph {et~al.}}]{Anderson:2012sa}%
  \BibitemOpen
  \bibfield  {author} {\bibinfo {author} {\bibfnamefont {L.}~\bibnamefont
  {Anderson}}, \bibinfo {author} {\bibfnamefont {E.}~\bibnamefont {Aubourg}},
  \bibinfo {author} {\bibfnamefont {S.}~\bibnamefont {Bailey}}, \bibinfo
  {author} {\bibfnamefont {D.}~\bibnamefont {Bizyaev}}, \bibinfo {author}
  {\bibfnamefont {M.}~\bibnamefont {Blanton}},  \emph {et~al.},\ }\href
  {\doibase 10.1093/mnras/sts084} {\bibfield  {journal} {\bibinfo  {journal}
  {Mon.Not.Roy.Astron.Soc.}\ }\textbf {\bibinfo {volume} {428}},\ \bibinfo
  {pages} {1036} (\bibinfo {year} {2013})},\ \Eprint
  {http://arxiv.org/abs/1203.6594} {arXiv:1203.6594 [astro-ph.CO]} \BibitemShut
  {NoStop}%
\bibitem [{\citenamefont {Padmanabhan}\ \emph {et~al.}(2012)\citenamefont
  {Padmanabhan}, \citenamefont {Xu}, \citenamefont {Eisenstein}, \citenamefont
  {Scalzo}, \citenamefont {Cuesta} \emph {et~al.}}]{Padmanabhan:2012hf}%
  \BibitemOpen
  \bibfield  {author} {\bibinfo {author} {\bibfnamefont {N.}~\bibnamefont
  {Padmanabhan}}, \bibinfo {author} {\bibfnamefont {X.}~\bibnamefont {Xu}},
  \bibinfo {author} {\bibfnamefont {D.~J.}\ \bibnamefont {Eisenstein}},
  \bibinfo {author} {\bibfnamefont {R.}~\bibnamefont {Scalzo}}, \bibinfo
  {author} {\bibfnamefont {A.~J.}\ \bibnamefont {Cuesta}},  \emph {et~al.},\
  }\href {\doibase 10.1111/j.1365-2966.2012.21888.x} {\bibfield  {journal}
  {\bibinfo  {journal} {Mon.Not.Roy.Astron.Soc.}\ }\textbf {\bibinfo {volume}
  {427}},\ \bibinfo {pages} {2132} (\bibinfo {year} {2012})},\ \Eprint
  {http://arxiv.org/abs/1202.0090} {arXiv:1202.0090 [astro-ph.CO]} \BibitemShut
  {NoStop}%
\bibitem [{\citenamefont {Blake}\ \emph {et~al.}(2011)\citenamefont {Blake},
  \citenamefont {Kazin}, \citenamefont {Beutler}, \citenamefont {Davis},
  \citenamefont {Parkinson} \emph {et~al.}}]{Blake:2011en}%
  \BibitemOpen
  \bibfield  {author} {\bibinfo {author} {\bibfnamefont {C.}~\bibnamefont
  {Blake}}, \bibinfo {author} {\bibfnamefont {E.}~\bibnamefont {Kazin}},
  \bibinfo {author} {\bibfnamefont {F.}~\bibnamefont {Beutler}}, \bibinfo
  {author} {\bibfnamefont {T.}~\bibnamefont {Davis}}, \bibinfo {author}
  {\bibfnamefont {D.}~\bibnamefont {Parkinson}},  \emph {et~al.},\ }\href
  {\doibase 10.1111/j.1365-2966.2011.19592.x} {\bibfield  {journal} {\bibinfo
  {journal} {Mon.Not.Roy.Astron.Soc.}\ }\textbf {\bibinfo {volume} {418}},\
  \bibinfo {pages} {1707} (\bibinfo {year} {2011})},\ \Eprint
  {http://arxiv.org/abs/1108.2635} {arXiv:1108.2635 [astro-ph.CO]} \BibitemShut
  {NoStop}%
\bibitem [{\citenamefont {Riess}\ \emph {et~al.}(2011)\citenamefont {Riess},
  \citenamefont {Macri}, \citenamefont {Casertano}, \citenamefont {Lampeitl},
  \citenamefont {Ferguson} \emph {et~al.}}]{Riess:2011yx}%
  \BibitemOpen
  \bibfield  {author} {\bibinfo {author} {\bibfnamefont {A.~G.}\ \bibnamefont
  {Riess}}, \bibinfo {author} {\bibfnamefont {L.}~\bibnamefont {Macri}},
  \bibinfo {author} {\bibfnamefont {S.}~\bibnamefont {Casertano}}, \bibinfo
  {author} {\bibfnamefont {H.}~\bibnamefont {Lampeitl}}, \bibinfo {author}
  {\bibfnamefont {H.~C.}\ \bibnamefont {Ferguson}},  \emph {et~al.},\ }\href
  {\doibase 10.1088/0004-637X/732/2/129, 10.1088/0004-637X/730/2/119}
  {\bibfield  {journal} {\bibinfo  {journal} {Astrophys.J.}\ }\textbf {\bibinfo
  {volume} {730}},\ \bibinfo {pages} {119} (\bibinfo {year} {2011})},\ \Eprint
  {http://arxiv.org/abs/1103.2976} {arXiv:1103.2976 [astro-ph.CO]} \BibitemShut
  {NoStop}%
\bibitem [{\citenamefont {Hu}\ and\ \citenamefont {White}(1997)}]{Hu:1997hp}%
  \BibitemOpen
  \bibfield  {author} {\bibinfo {author} {\bibfnamefont {W.}~\bibnamefont
  {Hu}}\ and\ \bibinfo {author} {\bibfnamefont {M.~J.}\ \bibnamefont {White}},\
  }\href {\doibase 10.1103/PhysRevD.56.596} {\bibfield  {journal} {\bibinfo
  {journal} {Phys.Rev.}\ }\textbf {\bibinfo {volume} {D56}},\ \bibinfo {pages}
  {596} (\bibinfo {year} {1997})},\ \Eprint
  {http://arxiv.org/abs/astro-ph/9702170} {arXiv:astro-ph/9702170 [astro-ph]}
  \BibitemShut {NoStop}%
\bibitem [{\citenamefont {Adshead}\ \emph {et~al.}(2012)\citenamefont
  {Adshead}, \citenamefont {Dvorkin}, \citenamefont {Hu},\ and\ \citenamefont
  {Lim}}]{Adshead:2011jq}%
  \BibitemOpen
  \bibfield  {author} {\bibinfo {author} {\bibfnamefont {P.}~\bibnamefont
  {Adshead}}, \bibinfo {author} {\bibfnamefont {C.}~\bibnamefont {Dvorkin}},
  \bibinfo {author} {\bibfnamefont {W.}~\bibnamefont {Hu}}, \ and\ \bibinfo
  {author} {\bibfnamefont {E.~A.}\ \bibnamefont {Lim}},\ }\href {\doibase
  10.1103/PhysRevD.85.023531} {\bibfield  {journal} {\bibinfo  {journal}
  {Phys.Rev.}\ }\textbf {\bibinfo {volume} {D85}},\ \bibinfo {pages} {023531}
  (\bibinfo {year} {2012})},\ \Eprint {http://arxiv.org/abs/1110.3050}
  {arXiv:1110.3050 [astro-ph.CO]} \BibitemShut {NoStop}%
\end{thebibliography}%
\vfill

\end{document}